\documentclass[sigconf]{acmart}
\acmSubmissionID{191}
\usepackage{graphicx}
\usepackage{amsmath}
\usepackage{subfig}
\usepackage{algorithm}
\usepackage{algorithmic}
\usepackage{amsthm}
\usepackage{multirow, booktabs}
\usepackage{subcaption}
\usepackage{hyperref}
\usepackage{mathrsfs}
\usepackage{multirow}
\usepackage{amsmath}
\usepackage{tabularx}
\usepackage{tabu}                     
\usepackage{multirow}                 
\usepackage{multicol}                 
\usepackage{float}                    
\usepackage{makecell}                 
\usepackage{amsmath, bm}
\usepackage{adjustbox}
\usepackage{ragged2e}
\usepackage{geometry}
\usepackage{fancyhdr}
\usepackage{caption}

\definecolor{hengpink}{cmyk}{0, 0.7808, 0.4429, 0.1412}

\usepackage{enumitem}

\newtheorem{proposition}{Proposition}
\begin{document}

\title{Hyperbolic Knowledge Transfer in Cross-Domain Recommendation System}
\author{Xin Yang}
\email{s2330128@u.tsukuba.ac.jp}
\affiliation{%
  \institution{University of Tsukuba}
  \country{Japan}
}
\author{Heng Chang}
\email{changh17@tsinghua.org.cn}
\affiliation{%
  \institution{Tsinghua University}
  \country{China}
}

\author{Zhijian Lai}
\email{galvin.lai@outlook.com}
\affiliation{%
  \institution{Peking University}
  \country{China}
}

\author{Jinze Yang}
\email{yangjinze@g.ecc.u-tokyo.ac.jp}
\affiliation{%
  \institution{University of Tokyo}
  \country{Japan}
}

\author{Xingrun Li}
\email{xr.li@amp.i.kyoto-u.ac.jp}
\affiliation{%
  \institution{Kyoto University}
  \country{Japan}
}

\author{Yu Lu}
\email{luyu06@baidu.com	}
\affiliation{%
  \institution{Baidu Inc.}
  \country{China}
}

\author{Shuaiqiang Wang}
\email{shqiang.wang@gmail.com}
\affiliation{%
  \institution{Baidu Inc.}
  \country{China}
}

\author{Dawei Yin}
\email{yindawei@acm.org}
\affiliation{%
  \institution{Baidu Inc.}
  \country{China}
}

\author{Erxue Min}
\authornote{Corresponding author.}
\email{erxue.min@gmail.com}
\affiliation{%
  \institution{Baidu Inc.}
  \country{China}
}
%
\begin{abstract}
Cross-Domain Recommendation (CDR) seeks to utilize knowledge from different domains to alleviate the problem of data sparsity in the target recommendation domain, and it has been gaining more attention in recent years. Although there have been notable advancements in this area, most current methods represent users and items in Euclidean space, which is not ideal for handling long-tail distributed data in recommendation systems. Additionally, adding data from other domains can worsen the long-tail characteristics of the entire dataset, making it harder to train CDR models effectively. Recent studies have shown that hyperbolic methods are particularly suitable for modeling long-tail distributions, which has led us to explore hyperbolic representations for users and items in CDR scenarios. However, due to the distinct characteristics of the different domains, applying hyperbolic representation learning to CDR tasks is quite challenging. In this paper, we introduce a new framework called Hyperbolic Contrastive Learning (HCTS), designed to capture the unique features of each domain while enabling efficient knowledge transfer between domains. We achieve this by embedding users and items from each domain separately and mapping them onto distinct hyperbolic manifolds with adjustable curvatures for prediction. To improve the representations of users and items in the target domain, we develop a hyperbolic contrastive learning module for knowledge transfer. Extensive experiments on real-world datasets demonstrate that hyperbolic manifolds are a promising alternative to Euclidean space for CDR tasks.

\end{abstract}

\ccsdesc[500]{Information systems~Recommender systems}

\keywords{Cross-domain Recommendation, Contrastive Learning, Hyperbolic Learning}
\maketitle
\section{Introduciton}
\justifying
To address the issue of information overload in our daily lives, recommender systems have been applied in numerous domains, such as e-commerce, video streaming platforms, and smartphone application markets \cite{taobao,youtube,w&d}.  The main idea of the recommender system is to take advantage of users' historical interaction data to infer their preferences. However, common recommender systems consistently encounter two challenges: 1) cold start issues, which arise when there are new users or items that the system has no adequate prior data; 2) the data sparsity problem\cite{cold}, which stems from the limited interactions between users and items.
\begin{figure}[t!]
    \vspace{2pt}
    \centering 
    \begin{minipage}[t]{0.23\textwidth}
        \vspace{0pt}
        \centering
        \includegraphics[width=1\linewidth]{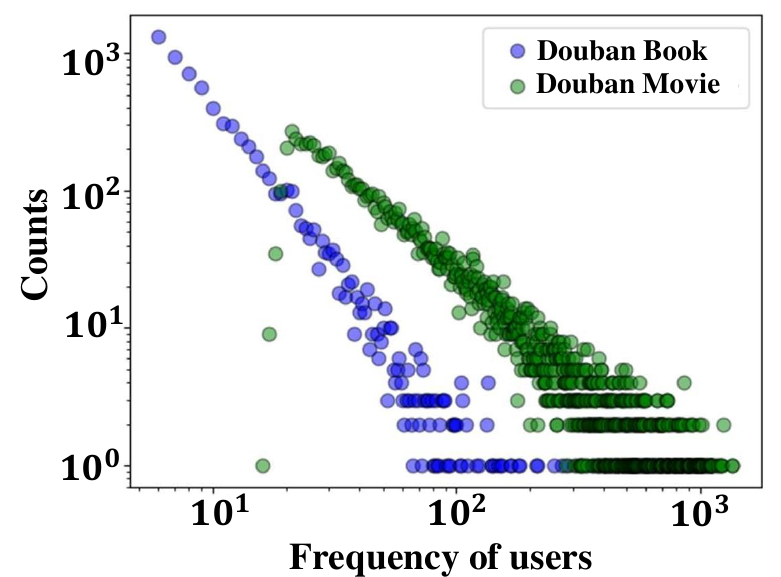}
        \captionsetup{font=footnotesize} 
\vspace{-3mm}
        
        \caption*{(a) Long-tail distribution in Douban Book and Douban Movie datasets.}
        
    \end{minipage}
   \hspace{0.1cm} 
    \begin{minipage}[t]{0.23\textwidth}
        \vspace{0pt}
        \centering
        \includegraphics[width=1\linewidth]{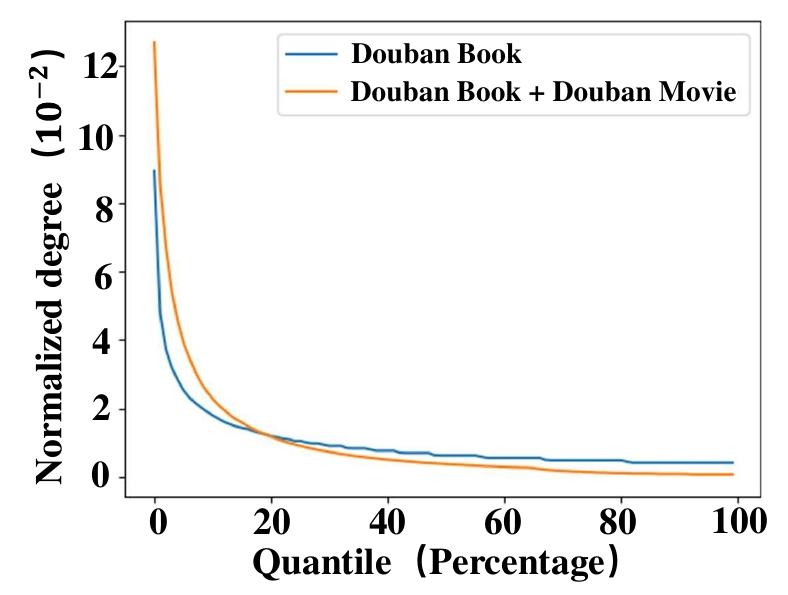}
        \captionsetup{font=footnotesize} 
\vspace{-3mm}
        \caption*{(b) A more significant long-tail trend after merging two datasets}
    \end{minipage}
\vspace{-3mm}
    \caption{ In (a), the x-axis represents the number of users per item, while the y-axis shows the count of associated items. In (b), the x-axis represents the intervals of the items. The y-axis represents the normalized degree.}
    \label{long_tail}
\vspace{-6mm}
\end{figure}

To address these challenges for a given recommendation scenario, a natural approach is to incorporate data from other related data sources or scenarios as supplements. Following this idea, Cross-Domain Recommendation (CDR) has attracted much attention in recent years, which aims to use the data of the source domain to enhance the model's performance on the target domain \cite{cmf,clfm,emcdr,biao,ccdr,conet,dtcdr} via various knowledge transfer strategies.

Despite the remarkable progress in CDR \cite{biao,ccdr,bitgcf,dtcdr,conet,cmf,emcdr,clfm,sscdr,coarse} in recent years, the long-tailed issue remains a major challenge in this field.
As illustrated in Figure \ref{long_tail} (a), in the book and movie domains of the Douban dataset, only a small fraction of items win the favor of a large number of user accounts, while the majority remain largely unpopular.
Worse still, in the task of CDR, merging data from different domains often exacerbates the long-tail distribution. 
For example, in Figure~\ref{long_tail} (b), we compare the difference in long-tail distributions before and after merging data from two domains. Specifically, for each dataset, we first sort users in terms of their degrees (the count of associated items) in decreasing order, 
then normalized the degrees by dividing total interaction counts for fair comparison. The merged dataset clearly exhibits a more pronounced long-tail distribution. This observation indicates that introducing data from other domains might exacerbate the long-tail nature of the entire dataset, making the learning of CDR models more challenging than traditional recommendation tasks.

One of the reasons why traditional neural network models perform poorly on data with a long-tail distribution is that they encode items and users into Euclidean space, which is a flat geometry with a polynomially expanding capacity. In fact, data with the long-tail distribution can be traced back to hierarchical structures \cite{hie}, whose number of items expanding exponentially. When encoding these data into Euclidean space, this imbalance makes it difficult to express the relationships among the data through the embedding vectors and subsequently deteriorates the accuracy of the final predictions.


In contrast, a hyperbolic manifold is a non-Euclidean space characterized by constant negative curvature, which allows the space to expand exponentially with the radius. This property of exponential expansion makes hyperbolic manifold particularly well-suited for representing tree-like or hierarchical structures, as it can naturally accommodate the exponential proliferation of elements in such structures \cite{hgcn,hnn,fhnn,h2h,hypercenter}.
For this reason, in recent years, significant advances have been made in hyperbolic neural networks to better handle the problem of long-tail distribution~\cite{hgcn,hnn,hgcf,fhnn,h2h,hgcc,hcl}.
 To accommodate graph-structured data, researchers have also proposed different hyperbolic graph neural network models \cite{hgcn,h2h,hgcf}.

Given the superior merits of hyperbolic representation learning in dealing with long-tail distribution data, in this paper, we focus on developing a hyperbolic neural network-based model for CDR tasks. However, the dissimilarities in characteristics between the two domains make it nontrivial to apply hyperbolic neural network-based methods to CDR tasks, which meets two challenges:
\begin{enumerate}[label=\textbf{C\arabic*},leftmargin=*]
    \item Source domain and target domain usually have different data distributions (e.g., user-item interactions are often richer in source domains while relatively sparse in target domains). How to effectively capture the inconsistency of different domains in the hyperbolic space is still unexplored.
    \item Although extensive earlier attempts have been proposed to transfer knowledge between the source domain and target domain in Euclidean space, it is still a challenging problem to transfer knowledge between two domains on a hyperbolic manifold.
\end{enumerate}

Considering the two challenges, we propose a novel framework named \textbf{\underline{H}}yperbolic \textbf{\underline{C}}on\textbf{\underline{T}}ra\textbf{\underline{S}}tive learning (\textbf{HCTS}) for the CDR tasks. 
To tackle the challenge \textbf{C1} and capture the domain specialties, we first use two GNN modules to perform neighbor propagation on the nodes in both domains separately and then embed users and items from the two domains onto two curvature-adaptive hyperbolic manifolds. The detached message-passing strategy ensures that the node embeddings are learned on the appropriate curvature for each domain.
Challenge \textbf{C2} arises from the inherent property of the hyperbolic manifold. 
Specifically, in Euclidean space, the curvature is constant 0 and the distance functions are uniform, making it straightforward to transfer knowledge across the source domain and the target domain \cite{ccdr,sscdr,emcdr}, 
while in the hyperbolic manifold, different curvatures can be defined but it is infeasible to compute the distance between two points located on hyperbolic manifolds with different curvatures\footnote{The hyperbolic distance function, as a well-defined function, which satisfies properties of positiveness, symmetry, and the triangle inequality, is only applicable for calculating distances within hyperbolic manifolds of uniform curvature.}.
To facilitate knowledge transfer across hyperbolic manifolds with inconsistent curvatures, we introduced a novel knowledge-transfer module based on hyperbolic contrastive learning, which is composed of three components: 1) \textbf{Manifold Alignment}: we leveraged the property that the tangent space at the north pole point is identical for hyperbolic manifolds of any curvature and defined a tailored projection layer, which aligns the embeddings from a hyperbolic manifold onto another; 2) \textbf{Hyperbolic Contrastive Learning}: 
 we propose three strategies of contrastive tasks, which are user-user, user-item, and item-item contrastive learning to transfer knowledge across domains from different aspects; 3) \textbf{Embedding Center Calibration}: We constrain the deviation of the geometric center of embeddings from the north pole point in the target domain to avoid the distortion of hyperbolic representations.

In summary, this work presents the following key contributions:
\begin{itemize}[leftmargin=*]
  \item To the best of our knowledge, we are the first to apply hyperbolic neural network-based models to CDR tasks and mitigate the long-tail distribution problem.
  \item We propose a novel hyperbolic contrastive learning framework that can effectively capture the distinctions of different domains and then transfer common knowledge across domains to facilitate the modeling of the target domain.
  \item We conducted extensive experiments on several datasets and compared our proposed methods with SOTA baselines. We proved that hyperbolic methods indeed serve as a compelling alternative to its Euclidean counterpart for the task of CDR.
\end{itemize}

\begin{figure*}[h!]
  \centering
  \includegraphics[width=0.9\textwidth]{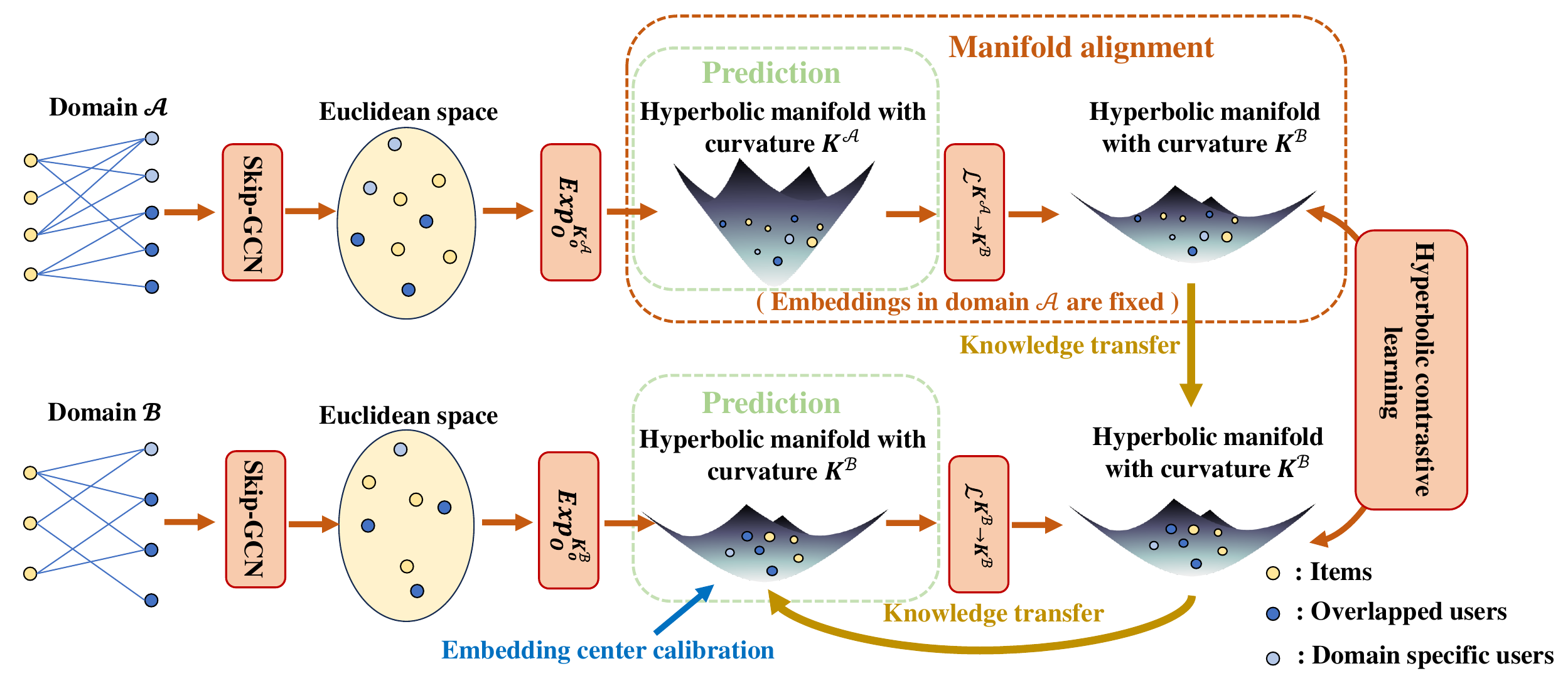}
\vspace{-6mm}
  \caption{The overall framework of HCTS. Graphs from the source and target domains are initially represented through independent embedding layers and skip-GCN layers to generate embeddings. Subsequently, these embeddings are mapped onto two distinct hyperbolic manifolds, each with an adaptive curvature, using an exponential map function for the source and target domains respectively. Predictions are made on these hyperbolic manifolds. To facilitate knowledge transfer from domain $\mathcal{A}$ to domain $\mathcal{B}$, the embeddings from one domain are transposed onto the hyperbolic manifold of the other domain using Equation \ref{ks-kt} and three contrastive learning tasks across domains are learned on the unified manifold. Additionally, a calibration task is designed to alleviate the embedding center deviation issue, avoiding the distortion of hyperbolic representations.}
  \label{overall}
\end{figure*}

\section{Preliminaries}
\subsection{Hyperbolic Geometry}
In this section, we introduce the fundamental concepts of hyperbolic geometry. More details are listed in Appendix~\ref{hyper}.

\textbf{Hyperbolic manifold.}
We define the Minkowski inner product
$
\langle\cdot,\cdot\rangle_{\mathcal{M}}: \mathbb{R}^{d+1} \times \mathbb{R}^{d+1} \rightarrow \mathbb{R}
$
by
\begin{equation*}
\langle x, y \rangle_{\mathcal{M}} := -x_0 y_0 + x_1 y_1 + \ldots + x_d y_d = x^\top J y,
\end{equation*}
where
$
J = \operatorname{diag}(-1,1,\ldots,1) \in \mathbb{R}^{(d+1) \times (d+1)}
$.
Given $K > 0$, we denote the $d$-dimensional hyperbolic manifold of constant negative curvature $-\frac{1}{K}$ as $\mathcal{H}^{K}$, defined as:
\begin{equation*}
\mathcal{H}^{K} := \{ x \in \mathbb{R}^{d+1} \,|\, \langle x, x \rangle_{\mathcal{M}} = -K, \, x_0 > 0 \}.
\end{equation*}
There is a special point called north pole point of $\mathcal{H}^{K} $:
\begin{equation*}
o:=(\sqrt{K},0,...,0) \in \mathcal{H}^{K}.
\end{equation*}

\textbf{Tangent space.} 
A tangent space centered at the point $x$ on $\mathcal{H}^{K}$ is denoted as $\mathcal{T}_x \mathcal{H}^{K}$, and is given as:
\begin{equation*}
\mathcal{T}_x \mathcal{H}^{K}= \{ v \in \mathbb{R}^{d+1} \, \big| \, \langle v, x \rangle_{\mathcal{M}} = 0 \}.
\end{equation*}
In particular, 
\begin{equation*}
\mathcal{T}_o \mathcal{H}^{K}=\{(0, x^{E} ) \in \mathbb{R}^{d+1} | x^{E} \in \mathbb{R}^d \} \cong \mathbb{R}^d,
\end{equation*}
where $x^{E}$ is an arbitrary $d$-dimensional Euclidean vector.
This advantage helps us to map the embeddings from one domain’s hyperbolic manifold to the other.

\textbf{Hyperbolic distance.} On a hyperbolic manifold, the distance (induced by Minkowski inner product) between two points $x$ and $y$ is given by 
\begin{equation}
d^K_{\mathcal{M}}=\sqrt{K}\text{arcosh}(-\langle x, y \rangle_\mathcal{M}/K).
\label{hyperbolic distance}
\end{equation}

\textbf{Exponential and logarithmic maps.} We can use an exponential map to map a point from a tangent space to a hyperbolic manifold and logarithmic maps to do the inverse. For $x,y \in \mathcal{H}^{d,K}$, $v\in \mathcal{T}_x \mathcal{H}^K$ such that $v \neq 0$ and $x\neq y$, the exponential and logarithmic maps of the hyperbolic manifold are given by
\begin{equation}\small
\mathcal{T}_x \mathcal{H}^K \ni \operatorname{Exp} ^K_x (v) = \cosh\left(\frac{\| v \|_{\mathcal{M}}}{\sqrt{K}}\right) \cdot x + \sqrt{K}\sinh\left(\frac{\| v \|_{\mathcal{M}}}{\sqrt{K}}\right)\cdot \frac{v}{\| v \|_{\mathcal{M}}}
\label{Exp}
\end{equation}
and
\begin{equation*}
\mathcal{H}^K \ni \operatorname{Log} ^K_x(y) = d^k_\mathcal{M}(x,y)\frac{y+\frac{1}{K}\langle x,y\rangle_\mathcal{M} \cdot x}{\|y+\frac{1}{K}\langle x,y\rangle_\mathcal{M} \cdot x\|}.
\end{equation*}

With exponential and logarithmic maps, it is possible to map embedding vectors from Euclidean space to hyperbolic space. For any $d$-dimensional vector $x^{E}$ in the Euclidean space, $(0,x^{E})$ is in the tangent space of the north pole point of $\mathcal{H}^{K}$. Subsequently, utilizing $\operatorname{Exp}_o^K$, we can map $x^{E}$ into hyperbolic space as follows:
\begin{equation}
x^{\mathcal{H}}:=\operatorname{Exp}_o^K((0,x^{\mathcal{E}})).
\label{e2h}
\end{equation}

\subsection{Cross-domain Recommendation}
A CDR task usually includes two domains: the source domain $\mathcal{S}=(\mathcal{U}^\mathcal{S},\mathcal{I}^\mathcal{S},\mathcal{E}^\mathcal{S})$ and the target domain $\mathcal{T}=(\mathcal{U}^\mathcal{T},\mathcal{I}^\mathcal{T},\mathcal{E}^\mathcal{T})$, where $\mathcal{U}$, $\mathcal{I}$ and $\mathcal{E}$ are the user set, the item set and the interactions between users and items. In this work, we only consider the situation where users overlap among the two domains. Formally, $\mathcal{U}^\mathcal{S}$ and $\mathcal{U}^\mathcal{T}$ can be redefined as $\mathcal{U}^\mathcal{S}=\mathcal{U}^{\mathcal{S}'}\cup\mathcal{U}^\mathcal{O}$ and $\mathcal{U}^\mathcal{T}=\mathcal{U}^{\mathcal{T}'}\cup\mathcal{U}^\mathcal{O}$, where $\mathcal{U}^{\mathcal{S}'}$ and $\mathcal{U}^{\mathcal{T}'}$ are the non-overlapping source user set and target user set, $\mathcal{U}^\mathcal{O}$ is the overlapping user set. The problem of the CDR task can be formally defined as follows.

\textbf{Input:} 
$\mathcal{S}=(\mathcal{U}^\mathcal{S},\mathcal{I}^\mathcal{S},\mathcal{E}^\mathcal{S})$,  $\mathcal{T}=(\mathcal{U}^\mathcal{T},\mathcal{I}^\mathcal{T},\mathcal{E}^\mathcal{T})$.

\textbf{Output:}
the learned function $\mathcal{F}=(u^\mathcal{T},i^\mathcal{T}|\mathcal{S},\mathcal{T},\Theta)$ that forecast whether $u^{\mathcal{T}}\in\mathcal{U}^\mathcal{T}$ would like to interact with $i^{\mathcal{T}}\in\mathcal{I}^\mathcal{T}$, where $\Theta$ denotes the model parameters.

\section{Methodology}
\subsection{Framework Overview}
In this section, we present the overview of our HCTS framework for CDR tasks. As depicted in Figure \ref{overall}, we first use an independent embedding layer and GNN layers to learn the embeddings of users and items in each domain separately, and then project the embeddings from two domains into two curvature-adaptive hyperbolic manifolds. The above practice enables the model to capture the specialty of each domain fully. After that, we employ a novel knowledge-transfer strategy that comprises three parts: 1) Manifold Alignment, to build an information bridge across manifolds of source and target domain; 2) Hyperbolic Contrastive Learning, to transfer knowledge across domains via three types of contrastive strategies; 3) Embedding Center Calibration, to mitigate the embedding center deviation issue caused by contrastive learning.  The contrastive learning losses, calibration loss, along hyperbolic margin ranking losses of both domains are jointly optimized to train the model. In the following sections, we introduce each part in detail.

\subsection{Embedding Layer}
All of $u^{\mathcal{S}}\in\mathcal{U}^{\mathcal{S}}$, $i^{\mathcal{S}}\in\mathcal{I}^{\mathcal{S}}$, $u^{\mathcal{T}}\in\mathcal{U}^{\mathcal{T}}$ and $i^{\mathcal{T}}\in\mathcal{I}^{\mathcal{T}}$ are fed into embedding layers independently, from which we obtain embeddings for source users $\mathbf{u}^{\mathcal{E},\mathcal{S}}$, source items $\mathbf{i}^{\mathcal{E},\mathcal{S}}$, target users $\mathbf{u}^{\mathcal{E},\mathcal{T}}$ and target items $\mathbf{i}^{\mathcal{E},\mathcal{T}}$. As we embed these users and items in Euclidean space, we denote them by a superscript $\mathcal{E}$. The overlapped users have embeddings in both the source domain and target domain, and non-overlapping users have only one embedding. When we do not need to distinguish $\mathcal{S}$ and $\mathcal{T}$, we simply write the embeddings of users and items as $\mathbf{u}^{\mathcal{E}}$ and $\mathbf{i}^{\mathcal{E}}$.

\subsection{Single Domain GNN Aggregator}

In most tasks, existing GNN models such as GCN \cite{gcn}, GraphSAGE \cite{sage}, GAT \cite{gat} or NGCF\cite{ngcf} are effective tools to perform message passing among nodes to enrich their embeddings with local structures. However, on the one hand,
recent studies in the field of recommender systems \cite{hgcf,lightgcn} have shown that the features of users and items in recommendation tasks are often highly sparse one-hot vectors, and therefore using feature transformation and activation functions does not necessarily improve the results. On the other hand, stacking multiple GNN layers together to fully exploit higher-order relations suffers from gradient vanishing or over-smoothing \cite{dropedge}.
For these two reasons, we use skip-GCN \cite{hgcf} for neighbor aggregation as follows:
\begin{align*}\textstyle
\mathbf{u}^{\mathcal{E},(l+1)} &= \mathbf{u}^{\mathcal{E},(l)}+\sum_{i\in N_u} \frac{1}{|N_u|}\mathbf{i}^{\mathcal{E},(l)}, \\
\mathbf{i}^{\mathcal{E},(l+1)} &= \mathbf{i}^{\mathcal{E},(l)}+\sum_{u\in N_i} \frac{1}{|N_i|}\mathbf{u}^{\mathcal{E},(l)}, 
\end{align*}
where the superscript $(l)$ means the layer of skip-GCN and the output of skip-GCN is
\begin{equation*}\textstyle
\tilde{\mathbf{u}}^{\mathcal{E}} = \sum_{k=1}^{L}\mathbf{u}^{\mathcal{E},(k)}, \qquad \tilde{\mathbf{i}}^{\mathcal{E}} = \sum_{k=1}^{L}\mathbf{i}^{\mathcal{E},(k)}.
\end{equation*}
We perform the neighbor aggregation on the source domain and target domain respectively and the final embedding of each user and item can be denoted as $\tilde{\mathbf{u}}^{\mathcal{E},\mathcal{S}}$, $\tilde{\mathbf{i}}^{\mathcal{E},\mathcal{S}}$ for the source domain, and $\tilde{\mathbf{u}}^{\mathcal{E},\mathcal{T}}$, $\tilde{\mathbf{i}}^{\mathcal{E},\mathcal{T}}$ for the target domain.

\subsection{Hyperbolic Manifold Projection}
In CDR tasks, the interaction graph from the target domain is often sparse and plagued with cold-start issues, whereas the interaction graph from the source domain is denser. Consequently, when performing downstream prediction tasks, the optimal curvature for the hyperbolic manifold differs between the source domain and the target domain. Therefore, we consider a trainable curvature for each domain, allowing both interaction graphs to be embedded into hyperbolic manifolds with their respective optimal curvatures. Then we use the exponential map function in Equation (\ref{e2h}) to map these embeddings to hyperbolic manifold as follows:
\begin{align*}
\mathbf{u}^{\mathcal{H}^{K_\mathcal{S}},\mathcal{S}} &= \operatorname{Exp}^{K_\mathcal{S}}_o(0,\tilde{\mathbf{u}}^{\mathcal{E},\mathcal{S}}), & \mathbf{i}^{\mathcal{H}^{K_\mathcal{S}},\mathcal{S}} &= \operatorname{Exp}^{K_\mathcal{S}}_o(0,\tilde{\mathbf{i}}^{\mathcal{E},\mathcal{S}}), \\
\mathbf{u}^{\mathcal{H}^{K_\mathcal{T}},\mathcal{T}} &= \operatorname{Exp}^{K_\mathcal{T}}_o(0,\tilde{\mathbf{u}}^{\mathcal{E},\mathcal{T}}), & \mathbf{i}^{\mathcal{H}^{K_\mathcal{T}},\mathcal{T}} &= \operatorname{Exp}^{K_\mathcal{T}}_o(0,\tilde{\mathbf{i}}^{\mathcal{E},\mathcal{T}}),
\end{align*}
where $K_{\mathcal{S}}$ and $K_{\mathcal{T}}$ are trainable parameters.

\subsection{Knowledge Transfer via Hyperbolic Contrastive Learning}
To transfer knowledge across node embeddings in hyperbolic manifolds with different curvatures, we proposed a novel hyperbolic contrastive learning framework. We elaborate on the details of each component in the following sections.

\textbf{Manifold Alignment.}
Since the embeddings in the source and target domains are projected onto hyperbolic manifolds with different curvatures, 
existing hyperbolic contrastive learning approaches \cite{hcts,hcts2,hcts3} cannot be directly applied because hyperbolic contrastive learning is based on hyperbolic distance and it is mathematically infeasible to compute the distance between points that are not located on the corresponding hyperbolic manifolds. To be specific, the hyperbolic distance function is derived from Minkowski inner product and is a well-defined function that satisfies properties of bi-linearity, positiveness, and symmetry, only within the tangent space of the corresponding hyperbolic manifold, as stated in Proposition~\ref{pro1}:
\begin{proposition}\label{pro1}
\(\langle \cdot, \cdot \rangle_{\mathcal{M}}\) is only a pseudo-inner product on \(\mathbb{R}^{n+1}\), however, it is an inner product restricted to the tangent spaces of \(\mathcal{H}^{n,K}\), i.e.,
\[
\langle \cdot, \cdot \rangle_{\mathcal{M}} : T_x\mathcal{H}^{n,K} \times T_x\mathcal{H}^{n,K} \rightarrow \mathbb{R}
\]
is a well-defined inner product on \( T_x\mathcal{H}^{n,K} \) for all \( x \in \mathcal{H}^{n,K} \). Then, \( \|u\|_{\mathcal{M}} = \sqrt{\langle u, u \rangle_{\mathcal{M}}} \) is a well-defined norm on it.
\end{proposition}
Proof for Proposition~\ref{pro1} can be found in Appendix \ref{PropositionProof}.
Accordingly, to perform hyperbolic contrastive learning between embeddings on hyperbolic manifolds with different curvatures, we have to first project the embeddings onto the hyperbolic manifold with a unified curvature. We define the linear transformation that maps the vectors from hyperbolic manifold with curvature $K_\mathcal{A}$ to the hyperbolic manifold with curvature $K_\mathcal{B}$ based on the fact that for different curvatures \( -\frac{1}{K} \), the manifolds \( \mathcal{H}^{n,K} \) have different north pole points but share the same tangent space at their north pole points. The linear transformation is defined as follows:
\begin{equation}
    \mathcal{L}^{K_\mathcal{A} \rightarrow K_\mathcal{B}}\left(x^{\mathcal{H}^{K_\mathcal{A}}}\right)=\operatorname{Exp} _o^{K_\mathcal{B}}\left(\operatorname{ReLU}\left(W \operatorname{Log} _o^{K_\mathcal{A}}\left(x^{\mathcal{H}^{K_\mathcal{A}}}\right)\right)\right).
    \label{h2h linear}
\end{equation}
To ensure $\operatorname{ReLU} \left(W \operatorname{Log} _o^{K_\mathcal{A}} x^{\mathcal{H}^{K_\mathcal{A}}}\right)$ stays in the tangent space of north pole points, we set $W$ as a trainable matrix with a fixed first row of zeros. The detailed explanation can be found in the remark of Appendix~\ref{e4}.

\begin{figure*}[h!]
    \centering 
    \begin{minipage}[b]{0.27\textwidth}
        \centering
        \includegraphics[width=\linewidth]{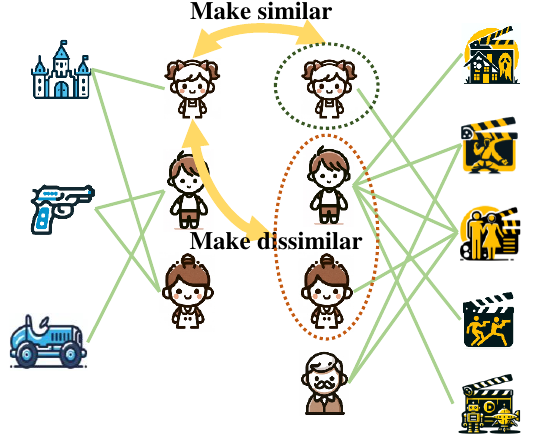}
    \vspace{-5mm}
        \caption*{(a) User-User contrastive learning}
        \label{u-u cts}
    \end{minipage}
    \hspace{0.5cm} 
    \begin{minipage}[b]{0.27\textwidth}
        \centering
        \includegraphics[width=\linewidth]{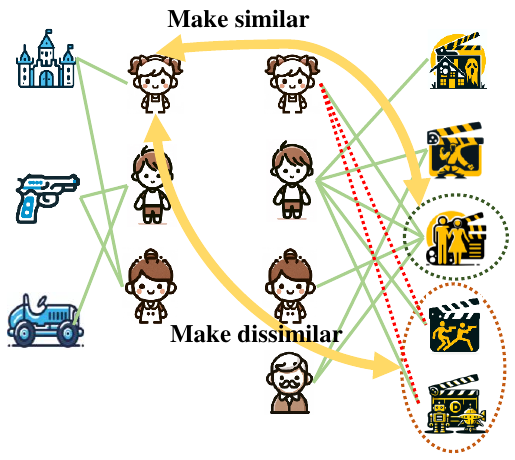}
    \vspace{-5mm}
        \caption*{(b) User-Item contrastive learning}
        \label{u-i cts}
    \end{minipage}
    \hspace{0.5cm} 
    \begin{minipage}[b]{0.27\textwidth}
        \centering
    \includegraphics[width=\linewidth]{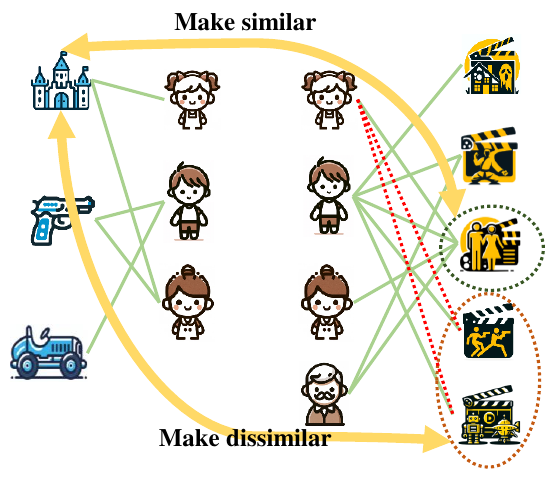}
    \vspace{-5mm}
        \caption*{(c) Item-Item contrastive learning}
        \label{i-i cts}
    \end{minipage}
    \vspace{-4mm}
    \caption{Three contrastive learning tasks across different domains.}
\label{strategy}
\vspace{-5mm}
\end{figure*}
\textbf{Hyperbolic Contrastive Learning.}
Based on the manifold alignment function provided above, we can employ contrastive learning techniques to transfer knowledge across two hyperbolic manifolds with different curvatures. To transfer knowledge from domain $\mathcal{A}$ to domain $\mathcal{B}$, where $\mathcal{A}$, $\mathcal{B}$ can be either $\mathcal{S}$ or $\mathcal{T}$, we first use the linear transformation in Equation (\ref{h2h linear}) to map the embeddings in domain $\mathcal{A}$ on hyperbolic manifold with curvature $K_\mathcal{A}$ onto the hyperbolic manifold with curvature $K_{\mathcal{B}}$ as follows:
\begin{align}
\mathbf{e}'^{\mathcal{H}^{K_\mathcal{B}},\mathcal{A}} &= \mathcal{L}^{{K_\mathcal{A}}\rightarrow{K_\mathcal{B}}}(\mathbf{e}^{\mathcal{H}^{K_\mathcal{A}},\mathcal{A}}),
\label{ks-kt}
\end{align} 
where $\mathbf{e}'^{\mathcal{H}^{K_\mathcal{B}},\mathcal{A}}$ denotes 
 the embedding from domain $\mathcal{A}$ on the hyperbolic manifold with curvature $K_\mathcal{B}$ and $\mathbf{e}$ can be either $\mathbf{u}$ or $\mathbf{i}$. In the process of transferring knowledge from domain $\mathcal{A}$ to domain $\mathcal{B}$, we keep the embeddings $\mathbf{e}'^{\mathcal{H}^{K_\mathcal{B}},\mathcal{A}}$ fixed (i.e., they are not updated during training).

For symmetry, we also implement linear transformation on embeddings in domain $\mathcal{B}$ on the same hyperbolic manifold:
\begin{align*}
\mathbf{e}'^{\mathcal{H}^{K_\mathcal{B}},\mathcal{B}} &= \mathcal{L}^{{K_\mathcal{B}}\rightarrow{K_\mathcal{B}}}(\mathbf{e}^{\mathcal{H}^{K_\mathcal{B}},\mathcal{B}}).
\end{align*} 
On a hyperbolic manifold with curvature $\mathcal{B}$, we defined three contrastive learning strategies, including user-user contrastive learning, user-item contrastive learning, and item-item contrastive learning, which are shown in Figure~\ref{strategy}.

We implement user-user contrastive learning to transfer knowledge among overlapped users. Although the behavior of a user may differ across two distinct domains, the correlation of behaviors of the same user across two domains is naturally higher than the correlation of behaviors from different users. Therefore, we have defined the following contrastive learning strategy:
\begin{align*}
L^{K_\mathcal{B}}_{u-u}=-\sum_{i\in\mathcal{U}^\mathcal{O}}\log\frac{\exp{(\operatorname{sim}(\mathbf{u}_i'^{\mathcal{H}^{K_\mathcal{B}},\mathcal{A}},\mathbf{u}_i'^{\mathcal{H}^{K_\mathcal{B}},\mathcal{B}})/\tau)}}{\sum_{j\in\mathcal{U}^\mathcal{O}\backslash{\{i\}}}\exp{(\operatorname{sim}(\mathbf{u}_i'^{\mathcal{H}^{K_\mathcal{B}},\mathcal{A}},\mathbf{u}_j'^{\mathcal{H}^{K_\mathcal{B}},\mathcal{B}})/\tau)}},
\end{align*}
where $\mathcal{U}^\mathcal{O}$ denotes the overlapped users. $\operatorname{sim}(\cdot,\cdot)$ is the similarity function defined as:
\begin{align*}
\operatorname{sim}(x,y)=-d^K_\mathcal{M}(x,y),
\end{align*}
where $d^K_\mathcal{M}$ is the hyperbolic distance on manifold with curvature $K$ and $\sigma$ is sigmoid function.
In this way, we ensure that the embeddings of the same user in different domains are closer, while those of different users in different domains become further.

We implement user-item contrastive learning for the consideration that when datasets from two domains are related, a user's behavior in one domain is highly related to that in another domain. For example, if a user likes action movies in a movie dataset, they would prefer gun toys or car toys over castle toys in the toy dataset. Therefore, we define user-item contrastive learning as follows: For overlapped users and the items in the other domain, we implement hyperbolic contrastive learning, which posits that user embeddings in one domain should closely align with the embeddings of items they interacted with in the other domain. In user-item contrastive learning, we sample one positive item and several negative items in domain $\mathcal{B}$.
\begin{align*}
L^{K_\mathcal{B}}_{u-i}=-\sum_{i\in\mathcal{U}^\mathcal{O}}\log\frac{\exp{(\operatorname{sim}(\mathbf{u}_i'^{\mathcal{H}^{K_\mathcal{B}},\mathcal{B}},\mathbf{i}_{pos}'^{\mathcal{H}^{K_\mathcal{B}},\mathcal{A}})/\tau)}}{\sum_{j\in N_{neg}}\exp{(\operatorname{sim}(\mathbf{u}_i'^{\mathcal{H}^{K_\mathcal{B}},\mathcal{B}},\mathbf{i}_j'^{\mathcal{H}^{K_\mathcal{B}},\mathcal{A}})/\tau)}},
\end{align*}
where ${i}_{pos}'$ denotes an item that interacted with user $i$ in the other domain. $N_{neg}$ denotes a set of items that did not interact with the user $i$ in the other domain. 
Similarly, to transfer knowledge between items interacted by the same user in different domains, we defined item-item contrastive learning as follows:
\begin{align*}
L^{K_\mathcal{B}}_{i-i}=-\sum_{i\in\mathcal{U}^\mathcal{O}}\log\frac{\exp{(\operatorname{sim}(\mathbf{i}_{pos}'^{\mathcal{H}^{K_\mathcal{B}},\mathcal{B}},\mathbf{i}_{pos}'^{\mathcal{H}^{K_\mathcal{B}},\mathcal{A}})/\tau)}}{\sum_{j\in N_{neg}}\exp{(\operatorname{sim}(\mathbf{i}_{pos}'^{\mathcal{H}^{K_\mathcal{B}},\mathcal{B}},\mathbf{i}_j'^{\mathcal{H}^{K_\mathcal{B}},\mathcal{A}})/\tau)}}.
\end{align*}
Based on the contrastive strategies discussed above, we first consider transferring knowledge from the source domain to the target domain via contrastive learning on the target manifold $K_\mathcal{T}$ and obtain three loss functions: $L^{K_\mathcal{T}}_{u-u}$, $L^{K_\mathcal{T}}_{u-i}$ and $L^{K_\mathcal{T}}_{i-i}$.
Additionally, since the quality of embeddings in the source domain determines the knowledge transferred to the target domain, we also consider the knowledge transfer from the target domain to the source domain and obtain $L^{K_\mathcal{S}}_{u-u}$, $L^{K_\mathcal{S}}_{u-i}$ and $L^{K_\mathcal{S}}_{i-i}$. In conclusion, the overall optimization objective of the contrastive knowledge transfer task is:
\begin{align*}
L_{cts}=L^{K_\mathcal{T}}_{u-u}+L^{K_\mathcal{T}}_{u-i}+L^{K_\mathcal{T}}_{i-i}+L^{K_\mathcal{S}}_{u-u}+L^{K_\mathcal{S}}_{u-i}+L^{K_\mathcal{S}}_{i-i}
\end{align*}

\begin{figure}[t!]
   \centering   \includegraphics[width=0.45\textwidth]{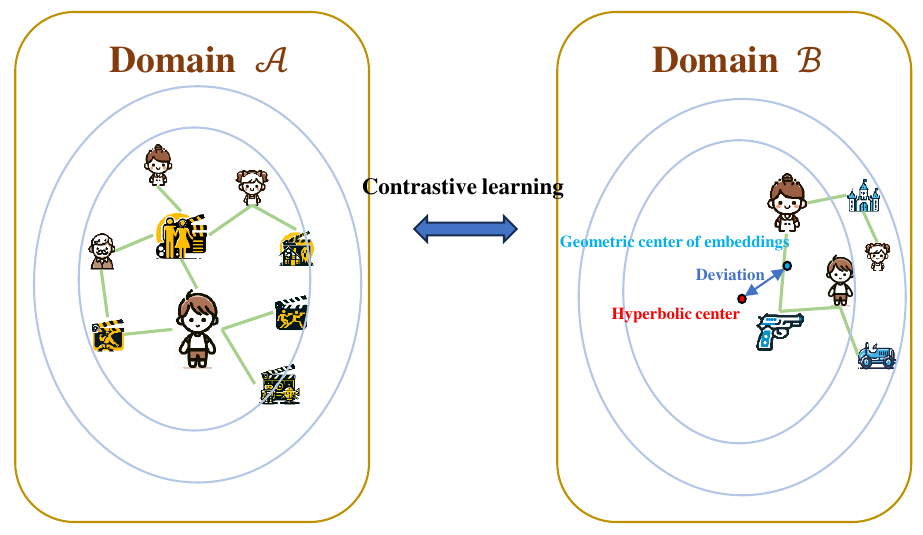}
   \vspace{-3mm}
   \caption{The deviation of embeddings from the north pole.
   }
   \vspace{-7mm}
   \label{fig:deviation}
 \end{figure}

\textbf{Embedding center calibration.}
Contrastive learning pushes the embeddings of similar nodes closer, whereas pulls the embeddings of dissimilar nodes further apart, thereby enhancing the discriminative nature of the learned representations. However, this process will lead to the deviation of the center of embeddings in the target domain from the north pole points of the hyperbolic manifold as depicted in Figure \ref{fig:deviation}, which might result in the distortion of hyperbolic representation and deteriorate its effectiveness in modeling the hierarchical structure of data.
Therefore, inspired by the investigation of the hyperbolic embedding center \cite{hypercenter}, we leverage a calibration function to correct the deviation of the embedding center. Specifically, we first map the hyperbolic embeddings back to the tangent space:
\begin{equation*}
\mathbf{e}^{\mathcal{E}}=\operatorname{Log} ^K_o(\mathbf{e}^{\mathcal{H}^K}),
\end{equation*}
where $\mathbf{e}$ are the vectors used for prediction and they can be both users and users. Then we calculate their geometric center as
\begin{equation*}
\mathbf{e}^{\mathcal{E}}_c = \frac{1}{|\mathcal{V}|} \sum_{i \in \mathcal{V}} \mathbf{e}^{\mathcal{E}}_{i},
\end{equation*}
and define the calibration loss
\begin{equation*}
 L_{clib}=\left(d^{\mathcal{E}}(\mathbf{e}^{\mathcal{E}}_c,o)\right)^2,
\end{equation*}
to enforce the geometric center of the embeddings close to the north pole point,
where $d^{\mathcal{E}}$ is Euclidean distance.

\subsection{Hyperbolic Margin Ranking Loss}
Margin ranking loss has been extensively used in recommendation tasks \cite{fgcn4rec,hgcf}, which separates positive and negative user-item pairs by a given margin. When the gap between a negative and a positive user-item pair exceeds this margin, neither pair contributes to the overall loss, which enables the optimization process to focus on the difficult pairs in the dataset. In this work, we use the hyperbolic version of margin ranking loss as the prediction loss. In the source domain, the prediction loss is:
\begin{equation*}
\begin{split}
L_{S}(u^{K_\mathcal{S}, \mathcal{S}}, i_{pos}^{K_\mathcal{S}, \mathcal{S}}, i_{neg}^{K_\mathcal{S}, \mathcal{S}}) = \max \bigg( d_{\mathcal{M}}(u^{K_\mathcal{S}, \mathcal{S}}, i_{pos}^{K_\mathcal{S}, \mathcal{S}})^2 \\
- d_{\mathcal{M}}(u^{K_\mathcal{S}, \mathcal{S}}, i_{neg}^{K_\mathcal{S}, \mathcal{S}})^2 + m, 0 \bigg),
\end{split}
\end{equation*}
where $m$ is a non-negative hyper-parameter, $u^{K_\mathcal{S}, \mathcal{S}}$  represents the embedding of users in the source domain on a hyperbolic manifold. $K_\mathcal{S}$ is derived by training to best fit the data from the source domain. $i_{pos}^{K_\mathcal{S}, \mathcal{S}}$ is the embedding of the positive sample of this user on this hyperbolic manifold, and $i_{neg}^{K_\mathcal{S}, \mathcal{S}}$ is embedding of a negative sample of this user on the same hyperbolic manifold. For the target domain, we get the prediction loss $L_{\mathcal{T}}$ in the same way.


\subsection{Multi-task Optimization}
We conduct a multi-task optimization for training the whole network. The loss function is defined as follows:
\begin{equation}
\begin{split}
L=L_{\mathcal{S}}+L_{\mathcal{T}}+\lambda_{cts} L_{cts}+\lambda_{clib} L_{clib},
\end{split}
\label{overall loss}
\end{equation}
where $\lambda_{cts}$ and $\lambda_{clib}$ are hyper-parameters ranging from 0 to 1.

\section{Experiments}
In this section, we conduct extensive experiments on multiple public datasets to evaluate the proposed method and primarily address the following questions:
\begin{description}
    \item[RQ1:] How does HCTS perform compared to baseline models on real-world datasets?
    \item[RQ2:] How does each proposed module contribute to the performance?
    \item[RQ3:] How does HCTS perform on the head and tail items?
    \item[RQ4:] How does HCTS influence the final embeddings of the source domain and target domain?
\end{description}
Extra experimental results and analysis are placed in Appendix~\ref{extra} due to the limit of pages.

\subsection{Experimental Setup}
\subsubsection{Datasets and Evaluation Protocols.}
We evaluated our proposed and baseline models on multiple real-world datasets, selecting four subsets from the Amazon dataset and three from the Douban dataset. Data statistics is detailed in Appendix~\ref{statistics}. We evaluate all these models by HR@10 and NDCG@10.
\subsubsection{Baselines.}
To demonstrate the effectiveness of our proposed model, we compare our model with three categories of models: (A) Single-domain GNN-based approaches, which include GCF\cite{bitgcf} and LightGCN\cite{lightgcn}. (B) Single-domain hyperbolic GNN-based method, which is HGCF\cite{hgcf} and (C) Cross-domain methods Bi-TGCF\cite{bitgcf}, CLFM\cite{clfm}, CMF\cite{cmf}, CoNet\cite{conet}, DTCDR\cite{dtcdr}, EMCDR\cite{emcdr}, CCDR\cite{ccdr} and ART-CAT\cite{artcat}. The description of these models is detailed in Appendix~\ref{baseline} and the implementation details are introduced in Appendix \ref{imp_details}.


\begin{table*}[ht]
\centering
\caption{Overview of performance. The left of $\rightarrow$ denotes the source domain dataset, and the name on the right denotes the target domain dataset. N@10 and H@10 are abbreviations for the metrics ndcg@10 and hit@10, respectively. * represents the significance level $p-$value $<0.05$. The highest scores for each dataset and metric are emphasized in bold, while the second-best ones are underlined. Improvement in the last line denotes the relative improvement compared to the best baseline.}
\label{overall results}
\vspace{-3mm}
\resizebox{\textwidth}{!}{
\begin{tabular}{lcccccccccccc}
\toprule
\multirow{2}{*}{Models} & \multicolumn{6}{c}{Amazon} & \multicolumn{6}{c}{Douban} \\
\cmidrule(lr){2-7} \cmidrule(lr){8-13}
 & \multicolumn{2}{c}{Book$\rightarrow$Movie} & \multicolumn{2}{c}{Book$\rightarrow$Music} & \multicolumn{2}{c}{Movie$\rightarrow$Toy} & \multicolumn{2}{c}{Book$\rightarrow$Music} & \multicolumn{2}{c}{Movie$\rightarrow$Book} & \multicolumn{2}{c}{Movie$\rightarrow$Music} \\
\cmidrule(lr){2-3} \cmidrule(lr){4-5} \cmidrule(lr){6-7} \cmidrule(lr){8-9} \cmidrule(lr){10-11} \cmidrule(lr){12-13}
& N@10 & H@10 & N@10 & H@10 & N@10 & H@10 & N@10 & H@10 & N@10 & H@10 & N@10 & H@10 \\
\midrule
HGCF & \underline{0.0347} & \underline{0.0942} & \underline{0.0488} & \underline{0.1163} & 0.0253 & 0.0568 & \underline{0.0444} & \underline{0.1722} & \underline{0.0481} & \underline{0.1990} & 0.0448 & 0.1722 \\
LightGCN & 0.0230 & 0.0653 & 0.0449 & 0.1066 & 0.0299 & 0.0591 & 0.0409 & 0.1643 & 0.0295 & 0.1260 & 0.0409 & 0.1643 \\
GCF & 0.0234 & 0.0668 & 0.0456 & 0.1076 & 0.0278 & 0.0544 & 0.0414 & 0.1626 & 0.0443 & 0.1718 & 0.0416 & 0.1661 \\
\hline
BiTGCF & 0.0271 & 0.0832 & 0.0460 & 0.1143 & \underline{0.0303} & \underline{0.0615} & 0.0403 & 0.1608 & 0.0430 & 0.1779 & 0.0451 &0.1775 \\
CoNet & 0.0131 & 0.0395 & 0.0102 & 0.0281 & 0.0095 & 0.0249 & 0.0212 & 0.0949 & 0.0373 & 0.1422 & 0.0216 & 0.0975 \\
DTCDR & 0.0132 & 0.0413 & 0.0231 & 0.0630  & 0.0150 & 0.0333 & 0.0207 & 0.0923 & 0.0406 & 0.1615 & 0.0276 & 0.1169 \\
CMF & 0.0241 & 0.0714 & 0.0421 & 0.1017 & 0.0289   & 0.0604 & 0.0384 & 0.1591 & 0.0428 & 0.1633 & 0.0430 & 0.1696 \\
DeepAPF & 0.0225 & 0.0649 & 0.0347 & 0.0930 & 0.0282 & 0.0568 & 0.0331 & 0.1371 & 0.0371 & 0.1555 & 0.0340 & 0.1371 \\
CLFM & 0.0157 & 0.0484 & 0.0253 & 0.0698 & 0.0161 & 0.0385 & 0.0215 & 0.1116 & 0.0345 & 0.1482 & 0.0258 & 0.1134 \\
EMCDR   & 0.0202 & 0.0573   & 0.0148 & 0.0453 & 0.0249 & 0.0568 & 0.0290 & 0.0833 & 0.0425 & 0.1045 & 0.0303 & 0.0833 \\
CCDR    & 0.0171 & 0.0557   & 0.0118 & 0.0397 & 0.0289 & 0.0605 & 0.0194 & 0.0448 & 0.0253 & 0.1125 & 0.0242 & 0.1142 \\
ART-CAT & 0.0236 & 0.0718&0.0334 &0.1008   & 0.0285 & 0.0586 &0.0308 & 0.1616 & 0.0462 & 0.1784 &\underline{0.0458} &\underline{0.1792} \\
\textbf{HCTS (ours)} & \textbf{0.0361*} & \textbf{0.0969*} & \textbf{0.0512*} & \textbf{0.1279*} & \textbf{0.0328*} & \textbf{0.0645*} & \textbf{0.0474*} & \textbf{0.1898*} & \textbf{0.0486*} & \textbf{0.2045*} & \textbf{0.0474*} & \textbf{0.1845*} \\
\hline
Improvement & +4.03\% & +2.86\% & +4.91\% & +9.97\% & +8.25\% & +4.87\% & +6.76\% & +10.22\% & +1.03\% & +2.76\% & +3.49\% & +2.95\% \\
\bottomrule
\end{tabular}
}
\end{table*}

\begin{table*}
\centering
\caption{Ablation study of our models on the datasets. 
\textit{HGCF-merge} denotes merging source domain dataset and target domain dataset and uses HGCF for prediction, H\textit{CTS-share} denotes HCTS with the same curvature of hyperbolic manifold for both source and target domains, HCTS-Euc denotes HCTS in Euclidean space, \textit{HCTS w/o s-t} denotes HCTS without transferring knowledge from the source domain to target domain, \textit{HCTS w/o t-s} denotes HCTS without transferring knowledge from the target domain to source domain, \textit{HCTS w/o center} denotes HCTS without center calibration, \textit{HCTS w/o u-u} denotes HCTS without user-user contrastive learning, \textit{HCTS w/o u-i} denotes HCTS without user-item contrastive learning,\textit{HCTS w/o i-i} denotes HCTS without item-item contrastive learning.}
\vspace{-3mm}
\normalsize 
\resizebox{\linewidth}{!}{
\begin{tabular}{lcc|cc|cc|cc|cc|cc}
\toprule
\multirow{4}{*}{Models} & \multicolumn{6}{c}{Amazon} & \multicolumn{6}{c}{Douban} \\
\cmidrule(lr){2-7} \cmidrule(lr){8-13}
& \multicolumn{2}{c}{Book$\rightarrow$Movie} & \multicolumn{2}{c}{Book$\rightarrow$Music} & \multicolumn{2}{c}{Movie$\rightarrow$Toy} & \multicolumn{2}{c}{Book$\rightarrow$Music} & \multicolumn{2}{c}{Movie$\rightarrow$Book} & \multicolumn{2}{c}{Movie$\rightarrow$Music} \\
\cmidrule(lr){2-3} \cmidrule(lr){4-5} \cmidrule(lr){6-7} \cmidrule(lr){8-9} \cmidrule(lr){10-11} \cmidrule(lr){12-13}
& N@10 & H@10 & N@10 & H@10 & N@10 & H@10 & N@10 & H@10 & N@10 & H@10 & N@10 & H@10\\
\midrule
    HGCF-merge & 0.0272 & 0.0752 & 0.0407 & 0.1095 & 0.0267 & 0.0591 & 0.0389 & 0.1503 & 0.0292 & 0.1355 & 0.0153 & 0.0835 \\
    HCTS-share      & 0.0318 & 0.0883 & 0.0509 & 0.1269 & 0.0293 & 0.0615 & 0.0463 & 0.1793 & 0.0452 & 0.1948 & 0.044  & 0.1714 \\
    HCTS-Euc  & 0.0216 & 0.0685  & 0.0404 & 0.1027   & 0.0228 & 0.0441 & 0.0328 & 0.1388 & 0.0475 & 0.1900 & 0.0457 & 0.1757 \\
    HCTS w/o s-t    & 0.0316 & 0.0888 & 0.0471 & 0.1134 & 0.0272 & 0.0591 & 0.0459 & 0.1754 & 0.0471 & 0.1963 & 0.0438 & 0.1722 \\
    HCTS w/o t-s    & 0.0353 & 0.0959 & 0.0479 & 0.1143 & 0.0279 & 0.0638 & 0.0472 & 0.1863 & 0.0461 & 0.1978 & 0.0473 & 0.1827 \\

    HCTS w/o u-u & 0.0345 & 0.0926 & 0.0503 & 0.1240  & 0.0257 & 0.0584 & 0.0471 & 0.1837 & 0.0450 & 0.1948 & 0.0475  & 0.1828 \\
    HCTS w/o u-i  & 0.0350 & 0.0954 & 0.0464 & 0.1114 & 0.0269 & 0.0612 & 0.0468 & 0.1837   & 0.0459 & 0.1990 & 0.0459 & 0.1775 \\
    HCTS w/o i-i      & 0.0356 & 0.0956  & 0.0389 & 0.1008     & 0.0271 & 0.0615 & 0.0472 & 0.1828   & 0.0462 & 0.1996 & 0.0476 & 0.1819 \\
    HCTS w/o center    & 0.0359 & 0.0967 & 0.0505 & 0.1261 & 0.0279 & 0.0631 & 0.0469 & 0.1801 & 0.0471 & 0.1984 & 0.0472 & 0.1837 \\
    \textbf{HCTS (ours)}       & \textbf{0.0361*} & \textbf{0.0969*} & \textbf{0.0512*} & \textbf{0.1279*} & \textbf{0.0328*} & \textbf{0.0645*} & \textbf{0.0474*} & \textbf{0.1898*} & \textbf{0.0486*} & \textbf{0.2045*} & \textbf{0.0474*} & \textbf{0.1845*} \\
\bottomrule
\label{ablation}
\end{tabular}
}
\vspace{-7mm}
\end{table*}

\begin{table*}
\centering
\caption{Performance on head 10\% and tail 90\% items on Douban Datasets. HCTS always performs the best for long-tail items.}
\vspace{-3mm}
\normalsize 
\resizebox{\linewidth}{!}{
\begin{tabular}{lcc|cc|cc|cc|cc|cc}
\toprule
\multirow{2}{*}{Models} & \multicolumn{4}{c}{DoubanMovie$\rightarrow$DoubanBook} & \multicolumn{4}{c}{DoubanMovie$\rightarrow$DoubanMusic} & \multicolumn{4}{c}{DoubanBook$\rightarrow$DoubanMusic} \\
\cmidrule(lr){2-5} \cmidrule(lr){6-9} \cmidrule(lr){10-13}
& N@10 head & N@10 tail & H@10 head & H@10 tail & N@10 head & N@10 tail & H@10 head & H@10 tail & N@10 head & N@10 tail & H@10 head & H@10 tail \\
\midrule
LightGCN  &0.0302             &0.0071             &0.1143             &0.0097                 &0.0355                &0.0064                      &0.1347          &0.0296            &\textbf{0.0355}  &0.0064   & 0.1347 & 0.0296 \\
HGCF      &0.0371  &\underline{0.011}             &0.1542             &\underline{0.0448}     &0.0346                &\underline{0.0102 }  &0.1309    &\underline{0.0413}&\underline{0.0346}&\underline{0.0102}   & 0.1309 & \underline{0.0413} \\
BiTGCF    &\textbf{0.0414}    &0.0016             &\textbf{0.1652}    &0.0127                 &\textbf{0.0370}       &0.0081                      &\textbf{0.1467}         &0.0308            &0.0306 &0.0097      &\underline{0.1422} & 0.0186 \\
\textbf{HCTS (ours)}      &\underline{0.0372} &\textbf{0.0114}    &\underline{0.1585} &\textbf{0.0460 }       &\underline{0.0368}    &\textbf{0.0106} &\underline{0.1423}   &\textbf{0.0422}   &0.0332        &\textbf{0.0142}    & \textbf{0.1485}    &\textbf{0.0413} \\
\bottomrule
\end{tabular}
}
\vspace{-3mm}
\label{tail}
\end{table*}



\subsection{Overall Performance Comparison (RQ1)}
In Table \ref{overall results}, the first three models represent single-domain approaches, which are trained on only the dataset of the target domain, and the subsequent ones are CDR models, which are trained on both the dataset of source and target domain. From the table, we can find the following observations:
\begin{enumerate}[label=(\arabic*),leftmargin=*]
    \item The hyperbolic representation learning-based models have remarkable superiority in dealing with recommendation tasks with long-tailed distribution. HGCF outperforms the best CDR baselines in Euclidean space in most cases and the most significant gap can be seen on Book-Movie of the Amazon dataset (NDCG@10: 0.0347 vs 0.0271, HIT@10: 0.0942 vs 0.0832). 
    \item HCTS outperforms HGCF significantly, it is because HCTS not only captures the hierarchical structure of data, but also effectively transfer knowledge from source domains to facilitate the learning on the target domain.
    \item HCTS outperforms all baselines but the improvement is also affected the correlation between the source domain and the target domain. On some datasets, the improvement is moderate, while in some others such as Book-Music tasks in both Amazon datasets and Douban datasets, the improvements are significant (+4.92\%, +5.09\% on NDCG@10 and +9.9\%, +10.2\% on HIT@10). This observation implies that the selection of auxiliary data sources is crucial.
    
\end{enumerate}

\subsection{Ablation Study (RQ2)}

To verify the effectiveness of each module, we conduct ablation studies over six public datasets. 

\begin{enumerate}[label=(\arabic*),leftmargin=*]
\item To determine the necessity of knowledge transfer between two domains, we merge the data from the source domain and the target domain and directly leverage HGCF for prediction. The performance, which is shown in the row HGCF-merge deteriorates a lot in all datasets and is even worse than models trained on merely the data from the target domain, in the HGCF row in Table\ref{overall results}. This is because the data distributions of the two domains are different and simply merging the data would result in the model dominated by the source domain, which contains richer interactions.

\item We evaluate the contribution of the trainable and independent curvatures by setting the curvatures of hyperbolic manifolds for both domains as the same value within HCTS. We can also find a remarkable performance decay in most datasets, which is shown in HGCF-share, such as on Book-Movie of the Amazon dataset (NDCG@10: 0.0361 vs 0.0318, HIT@10: 0.0969 vs 0.0883) and on Book-Music of the Douban dataset (NDCG@10: 0.0474 vs 0.0463, HIT@10: 0.1898 vs 0.1793).

\item To evaluate the contribution of hyperbolic manifold, we changed our model into a Euclidean version by removing the Exponential map and change the similarity function into cos similarity. We remain the graph convolution and contrastive learning strategy unchanged. We can find the results, which is shown in HCTS-Euc deteriorates remarkably in all datasets.

\item We also removed the hyperbolic contrastive learning tasks of source-to-target knowledge transfer and target-to-source knowledge transfer, which are denoted by HCTS w/o s-t and HCTS w/o t-s. And removed user-user, user-item, item-item contrastive learning strategy respectively, which are denoted by HCTS w/o u-u, HCTS w/o u-i and HCTS w/o i-i.

\item We removed the embedding center calibration module and also found a performance decline, especially on the Amazon Book-Movie and Movie-Toy datasets, which demonstrate the importance of removing the deviation.
\end{enumerate}

\subsection{Performance on Head and Tail items (RQ3)}
 To illustrate the validity of our proposal for long-tail distributions, we further conduct an in-depth analysis comparing the performance of head and tail items separately.
 Specifically, we divided the test data into head items, which refer to the portion of items that account for the top 10\% of popularity, and tail items, which refer to the remaining 90\% of items. We compared the results of two single-domain models and two cross-domain models with HCTS. It can be observed from Table~\ref{tail} that our model (HCTS) shows a significant improvement on tail items. Among the single-domain models, we can find that the performance of HGCF (hyperbolic representation-based model) on tail items is superior to Light-GCN (non-hyperbolic model). Similarly, in the cross-domain models, we find that our model (HCTS) also outperforms Bi-TGCF (non-hyperbolic model). Furthermore, we find that our model outperforms HGCF on both head and tail items, indicating that our model can effectively transfer useful knowledge from the source domain to the target domain, thereby improving performance across all items.
 
\begin{figure*}[h!]
    \centering 
    \begin{minipage}[b]{0.25\textwidth}
        \centering
        \includegraphics[width=\linewidth]{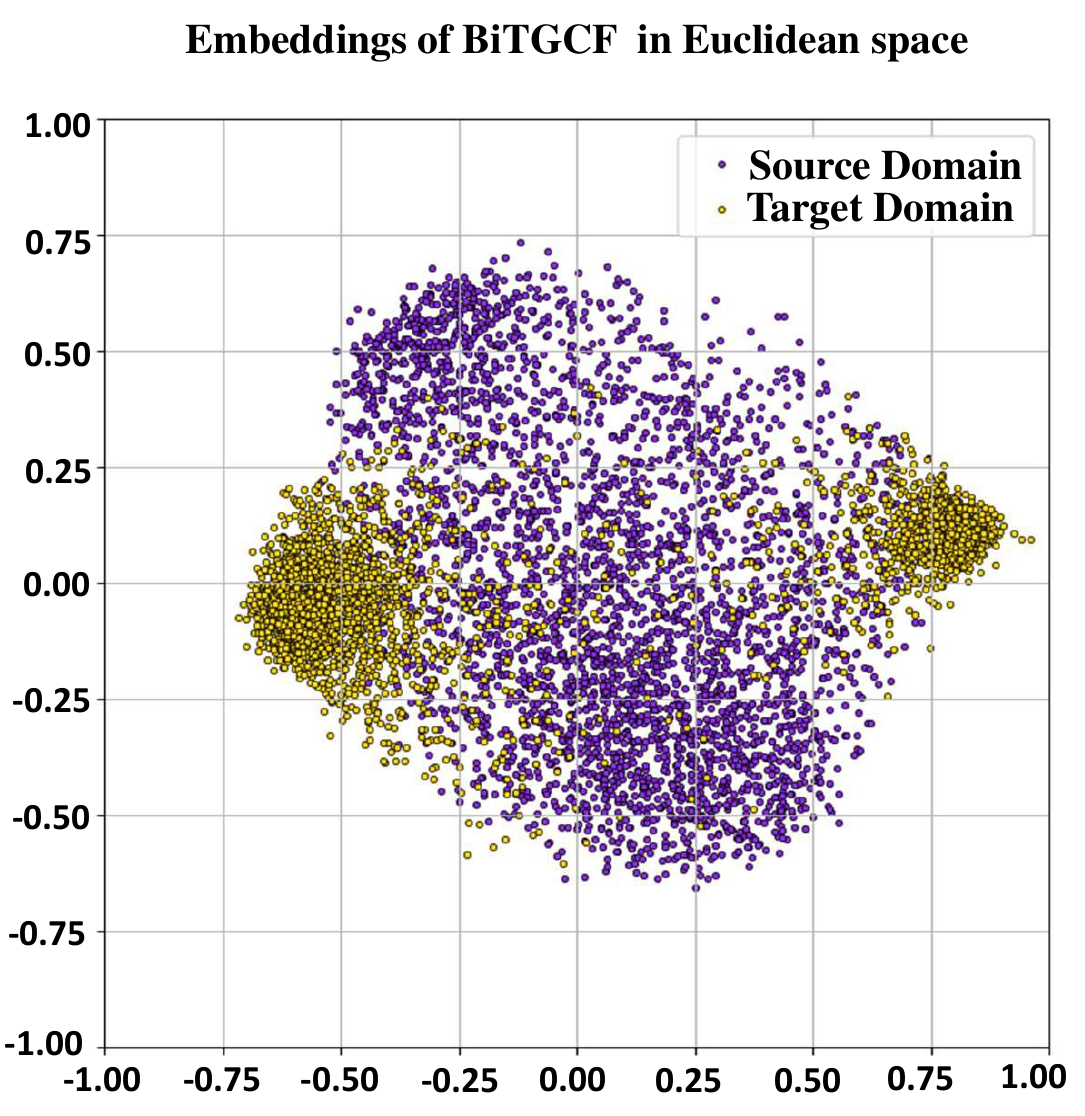}
        
    \end{minipage}
    \hspace{0.5cm} 
    \begin{minipage}[b]{0.25\textwidth}
        \centering
        \includegraphics[width=\linewidth]{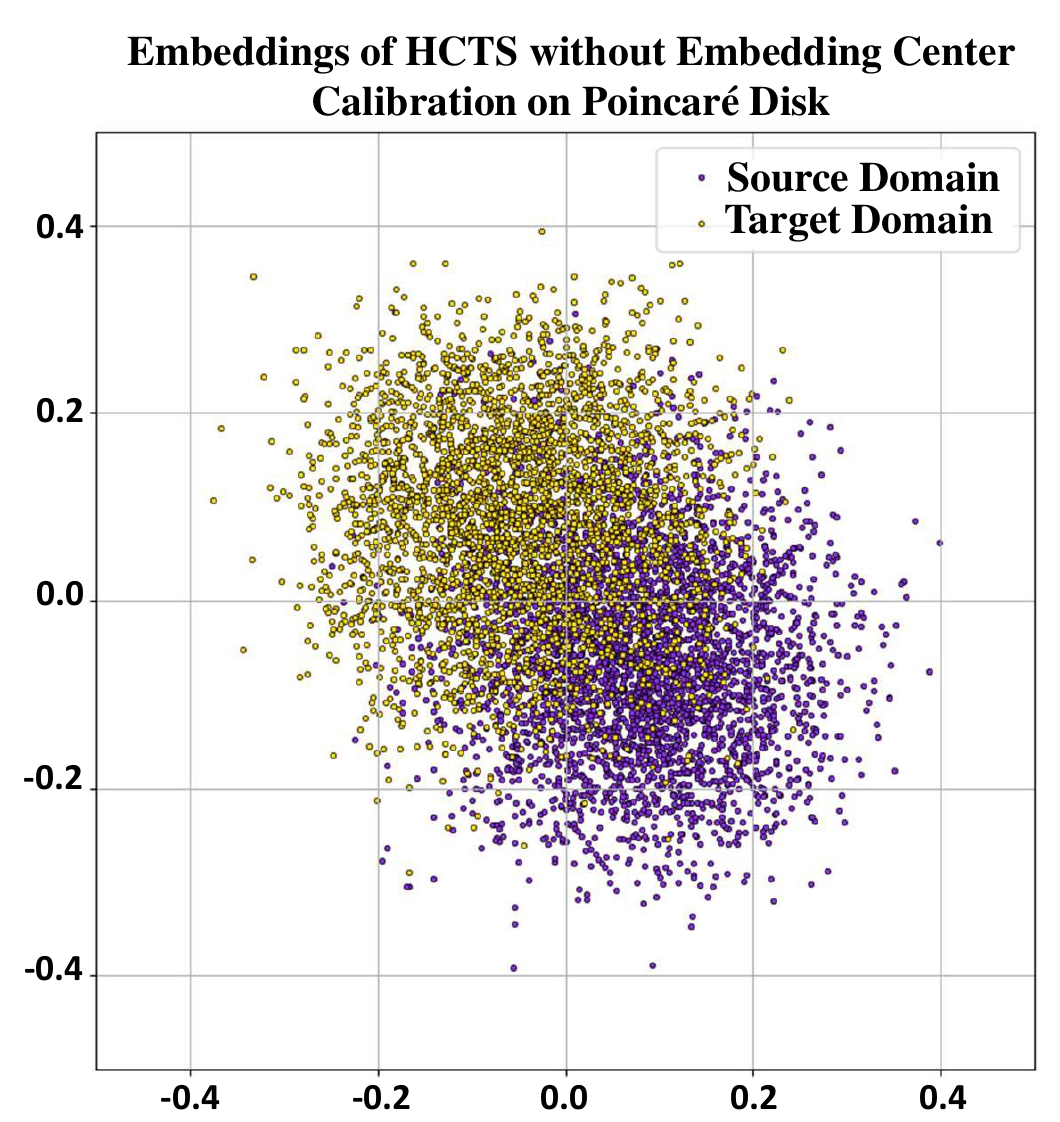}
    \end{minipage}
    \hspace{0.5cm} 
    \begin{minipage}[b]{0.25\textwidth}
        \centering
        \includegraphics[width=\linewidth]{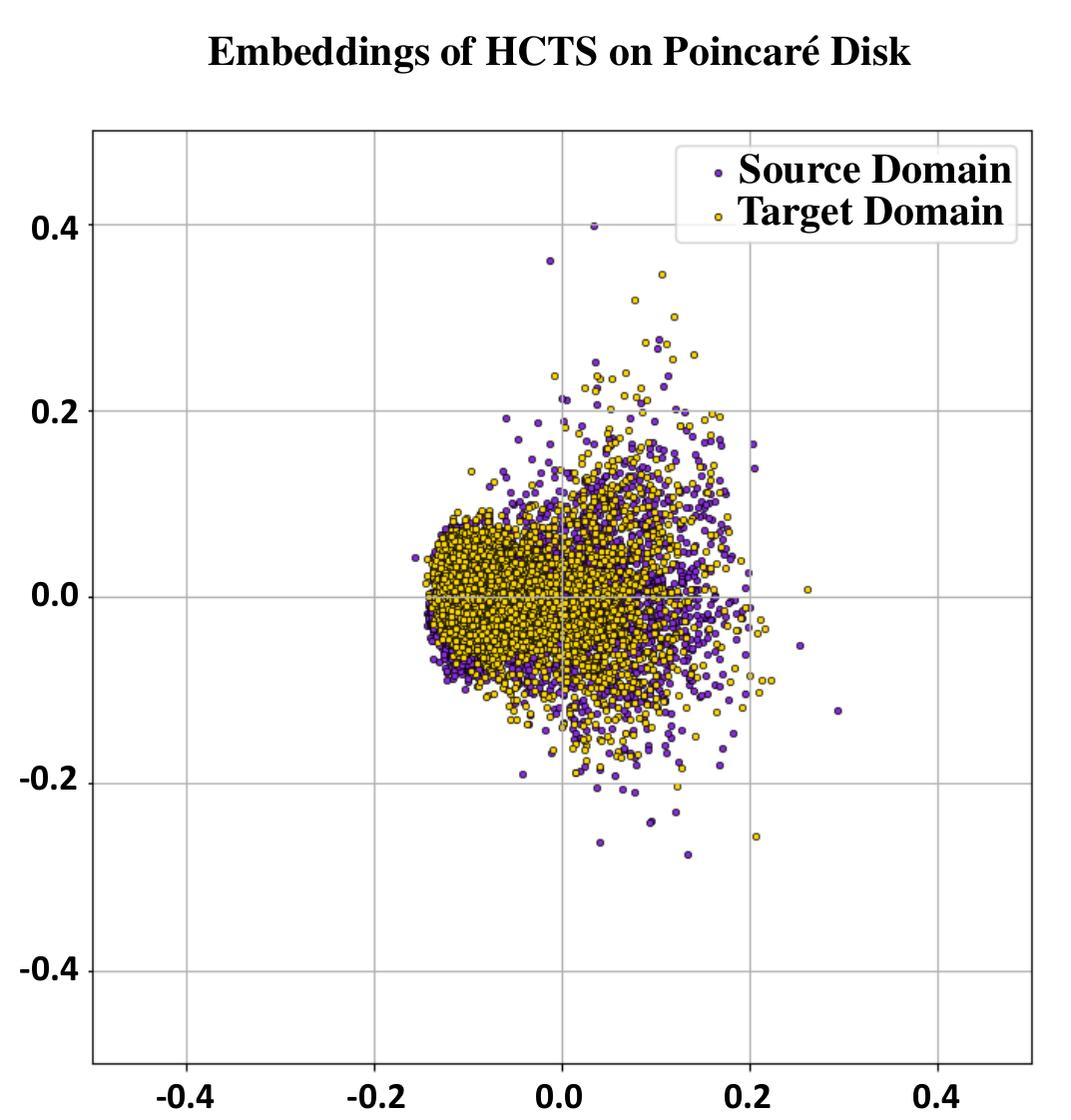}
    \end{minipage}
    \vspace{-3mm}
    \caption{Visualisation of item embeddings of BiTGCF in the two-dimensional Euclidean space (left), HCTS without embedding center calibration in the Poincaré representation of hyperbolic space (center) and HCTS without embedding center calibration in the Poincaré representation of hyperbolic space (right).}
    \vspace{-3mm}
\label{vision}
\end{figure*}
\subsection{Visualization of Embeddings (RQ4)}
To examine the impact of HCTS on the final embeddings of the source domain and target domain, we visualized the embeddings of HCTS, BiTGCF, and HCTS without embedding center calibration. For HCTS and HCTS without embedding center calibration, we visualized them using the Poincaré disk and projected the vectors to 2 dimensions. For BiTGCF, we directly reduced the dimension of vectors to 2. As shown in Figure~\ref{vision}, in BiTGCF (left), the embedding of the source domain is greatly influenced by the target domain during the knowledge transfer process. In contrast, HCTS (right) can transfer meaningful knowledge while ensuring that the original data information remains undamaged. 
Moreover, we find that without embedding center calibration, the geometric center of the embeddings will significantly deviate from the center of the hyperbolic manifold, as is shown in Figure~\ref{vision}(center). In contrast, our model, HCTS (right), successfully adjusts the geometric center of the embeddings back to the center of the hyperbolic manifold to better capture the hierarchical structure information of the data.

\section{Related works}
\textbf{Hyperbolic neural networks}. 
Due to the negative curvature characteristics of hyperbolic manifolds, they have advantages over Euclidean space when representing hierarchical structures, tree-like structures, and data with long-tail distributions. Researchers have invested significant effort in recent years to incorporate the geometric structure of hyperbolic manifolds into neural network models \cite{hnn,hgcn,hgcf,fhnn}. HNN \cite{hnn} first introduced the concept of hyperbolic neural networks, defining linear layers, activation functions, and recurrent neural networks in hyperbolic space. HGCN \cite{hgcn} extended graph neural networks to hyperbolic space, defining graph convolution in hyperbolic space. HGCF \cite{hgcf} was the first to use hyperbolic graph neural networks to address collaborative filtering problems, and since all its trainable parameters are defined in hyperbolic space, it employs the Riemannian optimizer for model training. FHNN \cite{fhnn} introduced the concept of the Lorentz linear layer, allowing the linear layers defined in hyperbolic space to be performed not in the tangent space, but entirely in hyperbolic space. The Hyperbolic-to-Hyperbolic Graph Convolutional Network \cite{h2h} uses the Einstein midpoint to approximate the average in hyperbolic space, defining a graph convolution operation that proceeds entirely in hyperbolic space without going through the tangent space. Hyperbolic contrastive learning \cite{hcts} proposed a method for contrastive learning in hyperbolic manifold and applied it to image recognition tasks and in the field of recommender system, \cite{hcts2} proposed a hyperbolic contrastive learning based method for sequential recommender system. In this study, to extend the CDR problem to hyperbolic manifolds and transfer knowledge across hyperbolic manifolds with different curvatures, we further introduce cross-domain hyperbolic contrastive learning for two bipartite graphs.

\textbf{Cross-domain recommendation}. 
CDR is a typical solution to solve cold start problems and data-sparse problems. The main idea is to transfer useful knowledge from the source domain to the target domain and improve the model's performance in the target domain. CLFM \cite{clfm}, which is constructed using a joint nonnegative matrix tri-factorization and proposed an effective alternating minimization strategy to ensure convergence. CMF \cite{cmf} is a method involving multi-relation learning that simultaneously processes the matrices of different domains by leveraging shared user latent factors. EMCDR \cite{emcdr}, which utilizes a latent factor model to learn user embeddings in the source domain and the target domain respectively. SSCDR \cite{sscdr} proposed a semi-supervised manner based on EMCDR, to learn item embeddings. Bi-TGCF \cite{bitgcf} is based on LightGCN \cite{lightgcn} and defined a transfer function that can enhance recommendation results in both domains. CCDR \cite{ccdr} is based on graph attention network \cite{gat} and employs contrastive learning to transfer knowledge. CoNet \cite{conet} is a deep model designed for cross-domain applications that facilitates knowledge transfer between domains through cross-connections within base networks. Based on CoNet, BIAO \cite{biao} proposed a behavior perceptron that predicts the importance of each source behavior. EMCDR \cite{emcdr} combines Matrix Factorization and Bayesian Personalized Ranking with a mapping function based on Multi-Layer Perceptron to align user latent factors across different domains. A multi-domain cross-domain recommendation model is proposed in\cite{artcat}, which applies contrastive autoencoder for learning single domain representations and an attention-based method for transferring knowledge. COAST\cite{coast} defined a cross-domain heterogeneous graph and a new message-passing mechanism to capture user-item interactions, and developed user-user and user-item interet invariance across domains to transfer knowledge. UniCDR\cite{unicdr} proposes a universal perspective for CDR. Despite the significant progress in CDR, as far as we know, the potential of hyperbolic representation learning has not been explored in this field.

\section{Conclusion}
In this study, we introduce a hyperbolic contrastive learning framework for cross-domain recommendation to fit the long-tail nature of data.  
Specifically, we embed data from two domains onto two hyperbolic manifolds with different learnable curvatures to capture the distribution discrepancy across domains. 
Then, we design a novel knowledge-transfer strategy based on hyperbolic contrastive learning to enrich the embeddings of the target domain and thus improve performance.
Our numerical experiments validate the effectiveness of our approach.

\newpage

\bibliographystyle{ACM-Reference-Format}

\bibliography{refs}

\appendix
\section*{Appendix}
\section{Hyperbolic geometry}\label{hyper}
In this appendix, we present details about the hyperbolic concepts mentioned in this work. 

\textbf{Minkowski (pseudo) inner product.} Consider the bilinear map 
$
\langle \cdot, \cdot \rangle_{\mathcal{M}} : \mathbb{R}^{n+1} \times \mathbb{R}^{n+1} \rightarrow \mathbb{R}
$
defined by
\[
\langle u, v \rangle_{\mathcal{M}} := -u_0v_0 + \sum_{i=1}^{n} u_iv_i = u^T J v,
\]
where \( J = \text{diag}(-1, 1, \ldots, 1) \in \mathbb{R}^{(n+1) \times (n+1)} \). It is called Minkowski (pseudo) inner product on \( \mathbb{R}^{n+1} \).
This is not an inner product on \(\mathbb{R}^{n+1}\) because \(J\) has a negative eigenvalue, but it is a pseudo-inner product because all eigenvalues of \(J\) are nonzero. Given a constant \(K > 0\), the equality
$
\langle x, x \rangle_{\mathcal{M}} = -K
$
implies that
\[
x_0^2 = K + \sum_{i=1}^{n} x_i^2 \geq K.
\]

\textbf{Hyperbolic manifold \(\mathcal{H}^{n,K}\) and Tangent space \( T_x\mathcal{H}^{n,K} \).}
Consider the subset of $\mathbb{R}^{n+1}$ defined as follows:
\begin{align*}
\mathcal{H}^{n,K} &:= \{ x \in \mathbb{R}^{n+1} : \langle x, x \rangle_{\mathcal{M}} = -K \text{ and } x_0 > 0 \} \\
&= \{ x \in \mathbb{R}^{n+1} : x_0^2 = K + x_1^2 + \ldots + x_n^2 \text{ and } x_0 > 0 \} \\
&= \{ x \in \mathbb{R}^{n+1} : h(x) := \langle x, x \rangle_{\mathcal{M}} + K = 0 \text{ with } x_0 > 0 \}.
\end{align*}
Here, the defining function \( h(x) = \langle x, x \rangle_{\mathcal{M}} + K \) has differential
\[
\mathrm{D}h(x)[u] = 2\langle x, u \rangle_{\mathcal{M}} = (2Jx)^{T} u. 
\]
Notice that \( x_0 \neq 0 \) for all \( x \in \mathcal{H}^{n,K} \). \( 2Jx \neq 0 \) for all \( x \in \mathcal{H}^{n,K} \) since \( J \) is invertible, and \( Jx = 0 \) if and only if \( x = 0 \). This implies that differential \( \mathrm{D}h(x) \colon \mathbb{R}^{n+1} \to \mathbb{R} \) is surjective (i.e., \(\text{rank } \mathrm{D}h(x) = 1\)) for all \( x \in \mathcal{H}^{n,K} \). By \cite[Definition 3.10 \& Theorem 3.15]{boumal2023intromanifolds}, we conclude the next definition.

\begin{definition}
    Given any constant \( K > 0 \), the set \( \mathcal{H}^{n,K} \) is an $n$-dimensional embedded submanifold of \( \mathbb{R}^{n+1} \) with tangent space:
    \begin{align*}
    T_x\mathcal{H}^{n,K} &= \ker \mathrm{D}h(x)
    = \{ u \in \mathbb{R}^{n+1} : \langle x, u \rangle_{\mathcal{M}} = 0 \} \nonumber \\
    &= \{ u \in \mathbb{R}^{n+1} : x_0u_0 = \sum_{i=1}^{n} x_iu_i \}
    \end{align*}
    which is an \( n \)-dimensional subspace of \( \mathbb{R}^{n+1} \).
\end{definition}

The restriction of \(\langle \cdot, \cdot \rangle_{\mathcal{M}}\) to each tangent space \(T_x\mathcal{H}^{n,K}\) defines a Riemannian metric on \(\mathcal{H}^{n,K}\), turning it into a Riemannian manifold. With this Riemannian structure, we call \(\mathcal{H}^{n,K}\) a \textit{hyperbolic manifold}. The main geometric trait of \(\mathcal{H}^{n,K}\) with \(n \geq 2\) is that its \textit{sectional curvatures} are negative constant, equal to
$
-\frac{1}{K}.
$

\textbf{North pole point on \(\mathcal{H}^{n,K}\).} The point \( o := (\sqrt{K}, 0, \ldots, 0) \in \mathcal{H}^{n,K} \) is called the north pole point of \( \mathcal{H}^{n,K} \).
We observe that
\begin{align*}
T_o\mathcal{H}^{n,K} &= \{ u \in \mathbb{R}^{n+1} : \langle o, u \rangle_\mathcal{M} = -\sqrt{K}, u_0 = 0 \} \\
&= \{ u \in \mathbb{R}^{n+1} : u_0 = 0 \} \\
&= \{ (0, u') \in \mathbb{R}^{n+1} : u' \in \mathbb{R}^n \} \\
&\cong \mathbb{R}^n.
\end{align*}
Thus, if we fix the dimension \( n \), then for different curvatures \( -\frac{1}{K} \), the manifolds \( \mathcal{H}^{n,K} \) have different north pole points but share the same tangent space at their north pole points. 


\textbf{Riemannian distance on \(\mathcal{H}^{n,K}\).} The distance function induced by Riemannian metric \(\langle \cdot, \cdot \rangle_{\mathcal{M}}\) is
\[
d^{K}_{\mathcal{M}}(x, y) = \sqrt{K} \operatorname{arcosh}\left(-\frac{\langle x, y \rangle_{\mathcal{M}}}{K}\right)
\]
for all \(x, y \in \mathcal{H}^{n,K}\).

\textbf{Exponential and logarithmic maps.} 
For $x \in \mathcal{H}^n_K, v \in T_x\mathcal{H}^n_K$ and $y \in \mathcal{H}^n_K$ such that $v \neq 0$ and $y \neq x$, the exponential and logarithmic maps are given by:
\[
\operatorname{Exp}_x^K(v) = \cosh\left(\frac{\|v\|_x}{\sqrt{K}}\right) \cdot x + \sqrt{K} \sinh\left(\frac{\|v\|_x}{\sqrt{K}}\right) \cdot \frac{v}{\|v\|_x}
\]
and
\begin{align*}
    \operatorname{Log}_x^K(y) &= \frac{d^K_\mathcal{M}(x, y)}{\|y + \frac{1}{K} \langle x, y \rangle_\mathcal{M} \cdot x\|_\mathcal{M}} \cdot \left( y + \frac{1}{K} \langle x, y \rangle_\mathcal{M} \cdot x \right) \\
    &= \frac{d^K_\mathcal{M}(x, y)}{\|\text{Proj}_x(y)\|_\mathcal{M}} \cdot \text{Proj}_x(y).
\end{align*}

\textbf{Example.} (Mapping from Euclidean space to hyperbolic manifold \cite{hgcn})

Let \( x^E \in \mathbb{R}^n \) denote input Euclidean features. Let \( o := (\sqrt{K}, 0, \ldots, 0) \) denote the north pole in \( \mathcal{H}^n_K \), which we use as a reference point to perform tangent space operations. We interpret \( (o, x^E) \) as a point in \( T_o\mathcal{H}^n_K \) and have
\begin{align*}
x^H &:= \exp_o^K ((o, x^E)) \\
&= \cosh\left(\frac{\|(o, x^E)\|_\mathcal{M}}{\sqrt{K}}\right) \cdot o + \sqrt{K} \sinh\left(\frac{\|(o, x^E)\|_\mathcal{M}}{\sqrt{K}}\right) \cdot \frac{(o, x^E)}{\|(o, x^E)\|_\mathcal{M}} \\
&= \cosh\left(\frac{\|x^E\|_2}{\sqrt{K}}\right) \cdot o + \sqrt{K} \sinh\left(\frac{\|x^E\|_2}{\sqrt{K}}\right) \cdot \frac{(0, x^E)}{\|x^E\|_2} \\
&= \left( \sqrt{K} \cosh\left(\frac{\|x^E\|_2}{\sqrt{K}}\right), \sqrt{K} \sinh\left(\frac{\|x^E\|_2}{\sqrt{K}}\right) \cdot \frac{x^E}{\|x^E\|_2} \right).
\end{align*}

For the last equality, notice the position of the zero elements in o and \( (0, x^E) \) as vectors of \( \mathbb{R}^n \).

\paragraph{Remark.}\label{e4} 
In the Equation~(\ref{h2h linear}), $\operatorname{Log} _o^{K_\mathcal{A}}\left(x^{\mathcal{H}^{K_\mathcal{A}}}\right)$ is in the tangent space of north pole point. We have to set the first row of $W$ to zero to ensure $W\operatorname{Log} _o^{K_\mathcal{A}}\left(x^{\mathcal{H}^{K_\mathcal{A}}}\right)$ is still in tangent space of north pole point. And because the north pole points of hyperbolic manifolds with different curvature share the same tangent space, it is possible to use Equation~\ref{h2h linear} to map vectors on the hyperbolic manifold with curvature $K_\mathcal{A}$ to the hyperbolic manifold with curvature $K_\mathcal{B}$.

\section{Proof of proposition~\ref{pro1}}
\label{PropositionProof}
\begin{proof}
Symmetry, additivity, and homogeneity hold since \(\langle u, v \rangle = u^T J v\) with diagonal \( J \). We next show positivity and definiteness. For all \(u\in T_x\mathcal{H}^{n,K}\), we have
\begin{align*}
\langle u, u \rangle_\mathcal{M} &= \left( \sum_{i=1}^{n} u_i^2 \right) - u_0^2 \\
&= \left( \sum_{i=1}^{n} u_i^2 \right) - \frac{1}{x_0^2}\left( \sum_{i=1}^{n} x_iu_i \right)^2 && \text{(by \( u \in T_x\mathcal{H}^{n,K} \),} \\
& && \text{\( \langle x, u \rangle_\mathcal{M} = 0 \) and \( x_0 > 0 \))} \\
&\geq \left( \sum_{i=1}^{n} u_i^2 \right) - \frac{1}{x_0^2}\left( \sum_{i=1}^{n} x_i^2 \right)\left( \sum_{i=1}^{n} u_i^2 \right) &&  \\
&= \left( \sum_{i=1}^{n} u_i^2 \right) \left( 1 - \frac{\sum_{i=1}^{n} x_i^2}{x_0^2} \right) \\
&= \left( \sum_{i=1}^{n} u_i^2 \right) \left( 1 - \frac{K}{x_0^2} \right) && \text{(by \( x \in \mathcal{H}^{n,K} \),} \\
& && \text{\( \langle x, x \rangle_\mathcal{M} = -K \))} \\
&= \frac{K}{x_0^2} \left( u_1^2 + \ldots + u_n^2 \right) > 0.
\end{align*}
Note that \( 0 < \frac{K}{x_0^2} \leq 1 \) here. If \(\langle u, u \rangle_\mathcal{M} = 0\), then by the above we have \(\sum_{i=1}^{n} u_i^2 = 0\); thus \(u_i = 0\) for \(i = 1, \ldots, n\). For \(i = 0\), \(u_0 = \frac{1}{x_0}\sum_{i=1}^{n} u_i^2 = 0\).
This completes the proof.
Remark that \(\langle u, u \rangle_\mathcal{M}\) can be negative if \(u\) does not belong to any tangent space of \(\mathcal{H}^{n,K}\).
\end{proof}

\section{Experimental Settings}
\subsection{Data Statistics}\label{statistics}
Table~\ref{datasets} shows the detailed statistics of the Amazon dataset\footnote{\url{http://snap.stanford.edu/data/amazon}} and three from the Douban dataset\footnote{\url{https://www.douban.com}} that we used in this research.
Table~\ref{overlap} shows the overlapping scale of different pairs of datasets. (Overlapping scale means the scale of overlapped users in the target dataset). As the overlapping scale of all the experiments on Douban datasets is 100\%, the overlapping scale is not shown in table \ref{overlap}.

\subsection{Details of Baseline Models}\label{baseline}
In this appendix, we introduce the baseline models in detail as follows:
\begin{itemize}[leftmargin=*]
\item[$\bullet$] \textbf{Bi-TGCF} \cite{bitgcf} is a GCN-integrated, dual-target CDR framework that simultaneously enhances recommendation results in both domains by facilitating a bidirectional knowledge transfer between them. 
\item[$\bullet$] \textbf{CLFM} \cite{clfm} is constructed using joint non-negative matrix tri-factorization and an effective alternating minimization strategy to ensure convergence.
\item[$\bullet$] \textbf{CMF} \cite{cmf} is a multirelation learning method that simultaneously processes the matrices of domains A and B by utilizing latent factors of the shared user. This approach starts with a joint learning process across the two domains and improves performance in the target domain.
\item[$\bullet$] \textbf{CoNet} \cite{conet} is a deep model designed for cross-domain applications that facilitates knowledge transfer between domains through cross-connections within base networks. This model undertakes a concurrent learning process across two domains and improves performance in both domains.
\item[$\bullet$] \textbf{DTCDR} \cite{dtcdr} is based on Multi-Task Learning (MTL), and an adaptable embedding sharing strategy to combine and share embeddings of common users across domains.
\item[$\bullet$] \textbf{GCF} \cite{bitgcf} is the version of Bi-TGCF without knowledge transfer between the source and target domain and it is a single domain recommendation model. 
\item[$\bullet$] \textbf{LightGCN} \cite{lightgcn} is a refined version of NGCF \cite{ngcf}, which is a graph neural network-based model for a single domain recommendation model. 
\item[$\bullet$] \textbf{HGCF} \cite{hgcf} is a hyperbolic graph neural network model for single-domain recommendation. 
\item[$\bullet$] \textbf{ART-CAT} \cite{artcat} is a multi-domain cross domain recommendation model, which applies Contrastive Autoencoder for learning single domain representations and Attention-based method for transferring knowledge. To align with our experimental settings, we only one dataset as source domain and one dataset as target domain.
\item[$\bullet$] \textbf{EMCDR} \cite{emcdr} combines Matrix Factorization and Bayesian Personalized Ranking with a mapping function based on Multi-Layer Perceptron to align user latent factors across different domains.
\item[$\bullet$] \textbf{CCDR} \cite{ccdr} is a GAT based model and use contrastive learning for knowledge transfer. 
\end{itemize}

\subsection{Implementation Details}
\label{imp_details}
All baseline models are implemented based on the famous recommendation system library Recbole
\cite{recbole,recbole2.0}. Specifically, Bi-TGCF, CLFM, CMF, CoNet, and DTCDR were directly adopted from Recbole CDR \cite{recbole2.0}'s implementations. The implementation of HGCF, CCDR and ART-CAT was based on their source code and was integrated into Recbole. We optimize all these methods except with Adam optimizer except HGCF, which is optimized by RSGD \cite{rsgd}. The learning rate of all these models is searched from $\{1e^{-2},1e^{-3},1e^{-4}\}$ and batch-size is searched from $\{1024,2048,4096\}$ and for fairness, all the embedding size is set as 64 and for CoNet, the MLP hidden size is set as [64,32,16,8]. 
To reduce the randomness in numerical experiments and make the results of each model more comparable, we computed full sort scores for every user, i.e., we calculated the scores for all items for each user. To maintain the settings of the CDR task mentioned in the Introduction, where the source domain dataset is a richer dataset and the target domain dataset is comparatively sparse, in the experiments below we use the dataset with a higher number of interactions as the source domain and the one with fewer interactions as the target domain.

\begin{figure*}[t]
  \centering
  \includegraphics[width=0.9\textwidth]{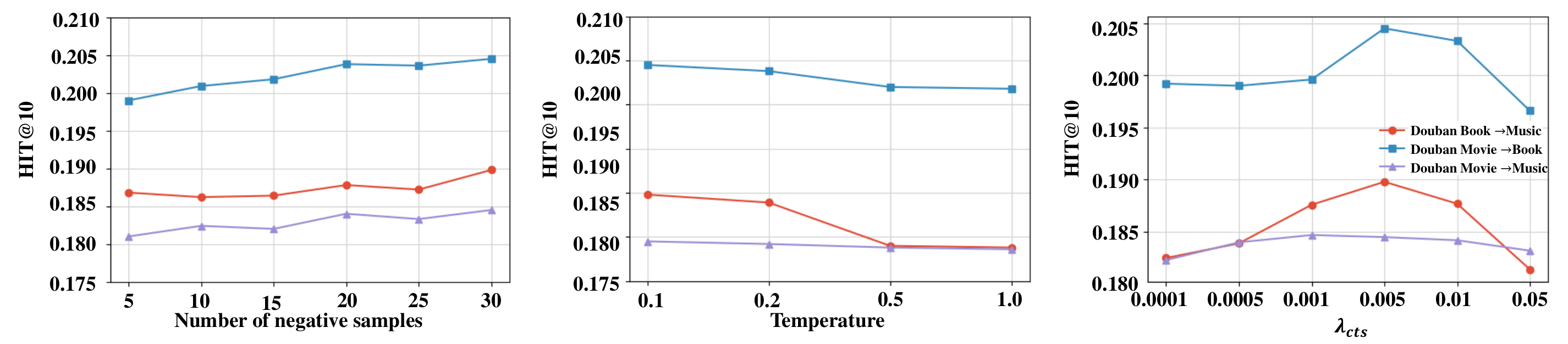}
  \caption{Sensitivity analysis on hyperparameters.}
  \label{exp:sensitivity}
\end{figure*}

\begin{table}[H]
\centering
\captionsetup{font=small, labelfont=bf}
\caption{Experimental datasets}
\label{datasets}
\setlength{\tabcolsep}{3pt} 
\renewcommand{\arraystretch}{0.9} 
\begin{tabular}{@{}lccccccc@{}}
\toprule
& \multicolumn{4}{c}{Amazon dataset} & \multicolumn{3}{c}{Douban dataset} \\
\cmidrule(lr){2-5} \cmidrule(lr){6-8}
& Book & Movie & Music & Toy & Book & Movie & Music \\
\midrule
Users & 5802 & 4616 & 1033 & 4264 & 1654 & 2620 & 1139 \\
Items & 21036 & 8762 & 2293 & 3503 & 6571 & 9522 & 5370 \\
Interactions & 284042 & 125889 & 18468 & 47570 & 88516 & 1017752 & 63312 \\
\bottomrule
\end{tabular}
\end{table}
\begin{table}[H]
\centering
\captionsetup{font=small, labelfont=bf}
\caption{Overlapping scale of different pairs of datasets.}
\label{overlap}
\setlength{\tabcolsep}{5pt} 
\begin{tabular}{@{}lc@{}}
\toprule
Experiments & Overlapping scale \\
\midrule
Book-Movie(Amazon) & 59.20\% \\
Book-Musics(Amazon) & 76.90\% \\
Movie-Musics(Amazon) & 74.30\% \\
Movie-Toy(Amazon) & 24.10\% \\
Toy-Movie(Amazon) & 22.09\% \\
Music-Movie(Amazon) & 16.50\% \\
Movie-Book(Amazon) & 47.52\% \\
Music-Book(Amazon) & 13.70\% \\
Toy-Book(Amazon) & 31.89\% \\
Music-toy(Amazon) & 6.67\% \\
Book-Toy(Amazon) & 43.39\% \\
\bottomrule
\end{tabular}
\end{table}

\section{Experimental Results}
\label{extra}
\subsection{Sensitivity Analysis}
We conducted sensitivity analysis on three hyperparameters of our model: number of negative samples for hyperbolic contrastive learning, temperature and $\lambda_{cts}$ And results are shown in Fig~\ref{exp:sensitivity}. We find that number of negative samples are stable and After the count exceeds 20, the value is almost constant. The optimal value for temperature is 0.1, attempting other values results in a gradual decline. And $\lambda_{cts}$ is also easy to find the optimal value is around 0.005 and 0.01.

\section{Computational costs of the HCTS framework}
\subsection{Time complexity of graph neural network-based baselines}
The followings are the notations used for analyzing time complexity.
\begin{itemize}
    \item Interactions in the source domain: $|E^S|$
    \item Interactions in the target domain: $|E^T|$
    \item Number of nodes in the source domain: $|S|$
    \item Number of nodes in the target domain: $|T|$
    \item Number of overlapped users: $|U^O|$
    \item Number of sampled users for contrastive learning: $B_1$
    \item Number of sampled items for contrastive learning: $B_2$
    \item Number of negative samples for contrastive learning: $N$
    \item Number of layers: $L$
    \item Latent dimension: $d$
    \item Users in testing set: $U_T$
    \item Items in testing set: $I_T$
\end{itemize}

\noindent Time complexity for each method:
\begin{itemize}
    \item \textbf{LightGCN}:
    \begin{itemize}
        \item Graph Encoding on target domain: $O(|E^T|d)$
        \item Inference : $O(|U_T||I_T|d)$
    \end{itemize}
    \item \textbf{GCF}:
    \begin{itemize}
        \item Graph Encoding on target domain: $O(|E^T|d)$
                \item Inference : $O(|U_T||I_T|d)$
    \end{itemize}
    \item \textbf{HGCF}:
    \begin{itemize}
        \item Exponential map: $O(|E^T|d)$
        \item Graph Encoding on target domain: $O(|E^T|d)$
        \item Inference : $O(|U_T||I_T|d)$
    \end{itemize}
    \item \textbf{BiTGCF}:
    \begin{itemize}
        \item Graph Encoding on target domain: $O(|E^T|d)$
        \item Graph Encoding on source domain: $O(|E^S|d)$
        \item Knowledge transfer: $O(|U^O|L)$
        \item Inference : $O(|U_T||I_T|d)$
    \end{itemize}
    \item \textbf{HCTS (ours)}:
    \begin{itemize}
        \item Exponential map: $O((|T| + |S|)d)$
        \item Graph Encoding on source domain: $O(|E^S|d)$
        \item Graph Encoding on target domain: $O(|E^T|d)$
        \item Knowledge transfer:
        \begin{itemize}
            \item Manifold alignment: $O((|T| + |S|)d)$
            \item User-user contrastive learning: $O(|U^O|d)$
            \item User-item contrastive learning: $O(B_1dN)$
            \item Item-item contrastive learning: $O(B_2dN)$
        \end{itemize}
        \item Inference : $O(|U_T||I_T|d)$
    \end{itemize}
\end{itemize}


\end{document}